\documentclass[pdflatex,sn-mathphys]{sn-jnl}

\usepackage{graphicx}%
\usepackage{multirow}%
\usepackage{amsmath,amssymb,amsfonts}%
\usepackage{amsthm}%
\usepackage{mathrsfs}%
\usepackage[title]{appendix}%
\usepackage{xcolor}%
\usepackage{textcomp}%
\usepackage{manyfoot}%
\usepackage{booktabs}%
\usepackage{algorithm}%
\usepackage{algorithmicx}%
\usepackage{algpseudocode}%
\usepackage{listings}%
\usepackage[sort&compress,numbers]{natbib}
\usepackage{xurl}

\usepackage{amsmath, amssymb, amscd, amsthm, amsfonts, graphicx, hyperref,array, xcolor, geometry, float, scalerel, booktabs, multirow, appendix,setspace, bm, empheq, lipsum,arydshln, placeins, anyfontsize, enumitem}

\jyear{2021}%

\theoremstyle{thmstyleone}%
\theoremstyle{thmstyletwo}%

\theoremstyle{thmstylethree}%

\begin{document}

\title[A quantitative model for melanoma cell population dynamics]{
\begin{center}
    A quantitative model for the emergent population dynamics of the melanoma MITF rheostat
\end{center}
}



\author*[1,2]{\fnm{Keith L.} \sur{Chambers}}\email{keith.chambers@ludwig.ox.ac.uk}

\author[1]{\fnm{Richard M.} \sur{White}}

\author[1]{\fnm{Colin R.} \sur{Goding}}

\author[1,2]{\fnm{Helen M.} \sur{Byrne}}

\affil[1]{\orgdiv{Ludwig Institute for Cancer Research}, \orgname{University of Oxford}, \orgaddress{\street{Old Road Campus Research Building, Roosevelt Dr, Headington}, \city{Oxford}, \postcode{OX3 7DQ}, \state{Oxfordshire}, \country{United Kingdom}}}

\affil[2]{\orgdiv{Wolfson Centre for Mathematical Biology}, \orgname{Mathematical Institute, University of Oxford}, \orgaddress{\street{Andrew Wiles Building, Radcliffe Observatory Quarter, Woodstock Road}, \city{Oxford}, \postcode{OX2 6GG}, \state{Oxfordshire}, \country{United Kingdom}}}

\abstract
{Cancer progression is driven by the ability of cells with identical driver mutations to adopt biologically distinct adaptive phenotypes. 
Yet the population dynamics implied by intratumour phenotypic heterogeneity is poorly understood. 
Melanoma is an excellent setting to study phenotype switching, in part because phenotypic identity is conferred by melanocyte inducing transcription factor (MITF) activity. 
Here we develop a multiscale phenotype-structured partial differential equation model for epidermal melanoma cell populations, first considering subcellular MITF and then {\color{black}{spatially uniform and spatially heterogeneous populations.}} 
The model admits three stable long-term  behaviours: slow growth with proliferative cells and non-cycling differentiated cells; faster expansion, with an invasive core; and rapid growth with oscillatory core dynamics. 
More broadly, the analysis highlights that phenotype reversibility by individual cells does not imply reversibility of phenotype population distributions. 
Hence, single-cell properties (e.g., reversibility of invasive capacity) must be extrapolated with caution to populations with coupled cell dynamics.
}







\maketitle

\clearpage

\section{Introduction} \label{sec: intro}

Cell plasticity is a widespread biological phenomenon that refers to the ability for cells to reversibly adapt their state, or \textit{phenotype}, in response to environmental cues \cite{merrell2016adult}. 
This adaptive capacity gives rise to phenotypic heterogeneity within cell populations that obscures understanding of many diseases \cite{mei2020plasticity, lin2024chronic, li2020cell}.
Importantly, cell plasticity is universal 
to cancers and considered a barrier to effective treatment due to its role in producing drug-tolerant tumour cells \cite{shenoy2020cell, marusyk2020intratumor}.
Several mathematical studies have been published concerning cell plasticity in cancer, including analysis of structured individual-based models \cite{stace2020discrete, browning2025identifiability}, stochastic branching processes \cite{GUNNARSSON2020110162, colson2025mathematical}, structured partial differential equation (PDE) models \cite{Clairambault2019AdaptiveSurvey, Cassidy2021} and multiscale models \cite{Anderson2006TumorMorphology}.

Melanoma is an excellent paradigm for cancer cell plasticity. 
It is typically considered a skin cancer, but may also arise in the eye and mucous membranes \cite{Caraviello2025MelanomaReview, Singh2018UvealMelanomaReview, Sergi2023MucosalMelanoma}.
{\color{black}Melanoma is characterised by the deregulated proliferation of melanocytes and, more broadly, cells in the melanocyte lineage derived from neural crest progenitors \cite{Caraviello2025MelanomaReview}. 
Melanocytes are pigment-producing cells found in the dermal-epidermal junction (the epidermis and dermis are the upper two layers of skin).}
Despite its lower incidence relative to other skin cancers, melanoma is disproportionately responsible for the majority of skin cancer-related deaths \cite{zhou2025melanoma_nmsc_burden}.
Its clinical importance has led to the establishment of a considerable number of mathematical and computational studies (reviewed in \cite{Albrecht2020ComputationalModelsMelanoma}).
Existing work includes ordinary differential equation (ODE) models for tumour-immune interactions and virotherapy  \cite{Rodrigues2025CARTMelanoma, Ramaj2023OncolyticMelanoma}, delay differential equation models for dendritic cell therapy \cite{DePillis2013DendriticMelanoma, CastilloMontiel2015DendriticCellMelanoma, Dickman2020DendriticCellTherapyMelanoma}, PDE models for combination therapies \cite{Lai2017CombinationTherapy, Nave2022MelanomaBRAFPD1} and agent-based models that capture tumour angiogenesis and genetic heterogeneity \cite{Wang2013MultiscaleMelanoma, Jamshad2026HallmarkIntegrated}. 
Importantly, several recent large-scale ODE models have also been developed to capture phenotypic plasticity via subcellular gene regulatory networks \cite{hari2022landscape, Subhadarshini2023PDL1Heterogeneity}, but these studies do not pursue explicit treatment of \textit{in vivo} population dynamics. 
A complementary study analysed a reduced gene regulatory network across an idealised, static Voronoi monolayer \cite{taylor2026travelling}.

\begin{figure}
    \centering
    \includegraphics[width=0.6\textwidth]{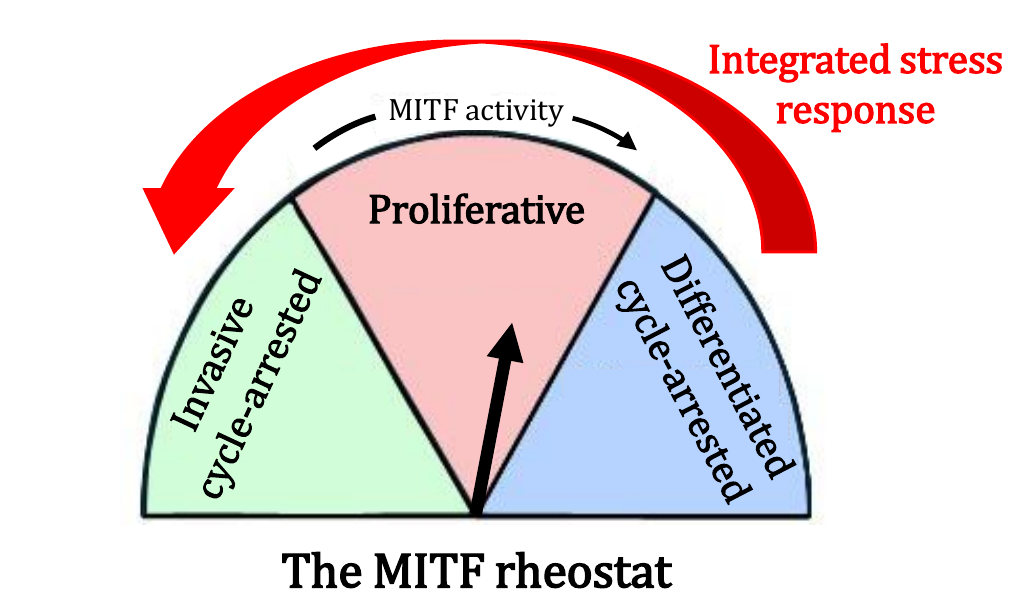}
    \caption{\textbf{The MITF rheostat associates melanoma cell phenotype with the activity of the MITF protein \cite{carreira2006mitf}. } Cells reversibly transition between three key states in a manner coupled to the integrated stress response. }
    \label{fig: rheostat}
\end{figure}

Melanoma is an ideal setting to study cancer cell plasticity due to the co-existence of multiple phenotypes within individual tumours \cite{tirosh2016dissecting, rambow2018toward, Goding2019MITFFirst25Years}. Moreover, a master regulator of phenotypic states has been identified and well-characterised, namely the {\color{black}\textit{melanocyte inducing transcription factor} (MITF; formerly known as microphthalmia-associated transcription factor) \cite{Goding2019MITFFirst25Years, Rambow2019MelanomaPlasticity}.}  
The relationship between MITF and melanoma cell phenotype is captured by the MITF rheostat model \cite{carreira2006mitf, Hoek2010PhenotypeSwitching}. 
This theory asserts that phenotype reversibly transitions between three key states in a manner driven by the activity of MITF protein. 
Here \textit{activity} refers to the downstream transcriptional and proteomic influence of MITF on cell state. 
High MITF activity gives rise to a differentiated and cycle-arrested state (DIF), intermediate activity promotes a proliferative state (PRO) and low activity generates an invasive cycle-arrested state (INV). 
Importantly, MITF is inversely coupled to the integrated stress response (ISR) \cite{PakosZebrucka2016ISR}; stressful microenvironments are associated with lower MITF expression that is at least partly attributable to elevated transcriptional repression via the ISR effector ATF4 and decreased MITF translation \cite{falletta2017translation}.
A summary is provided in Fig. \ref{fig: rheostat}.

The MITF rheostat is a theory of \textit{individual} cell state.
However, melanoma tumours comprise populations of malignant cells that collectively influence patient outcome via their coupled dynamics. 
This raises the research question considered in the present work: \\ \\
\textbf{Key question: }What are the emergent population dynamics implied by the MITF rheostat? More specifically,
\begin{itemize}
    \item[-] What is the phenotype distribution amongst melanoma cells?
    \item[-] What long-term qualitative behaviours emerge from the interplay between phenotypic plasticity, proliferation and cell death?  \\
\end{itemize}

\begin{figure}
    \centering
    \includegraphics[width=0.99\textwidth]{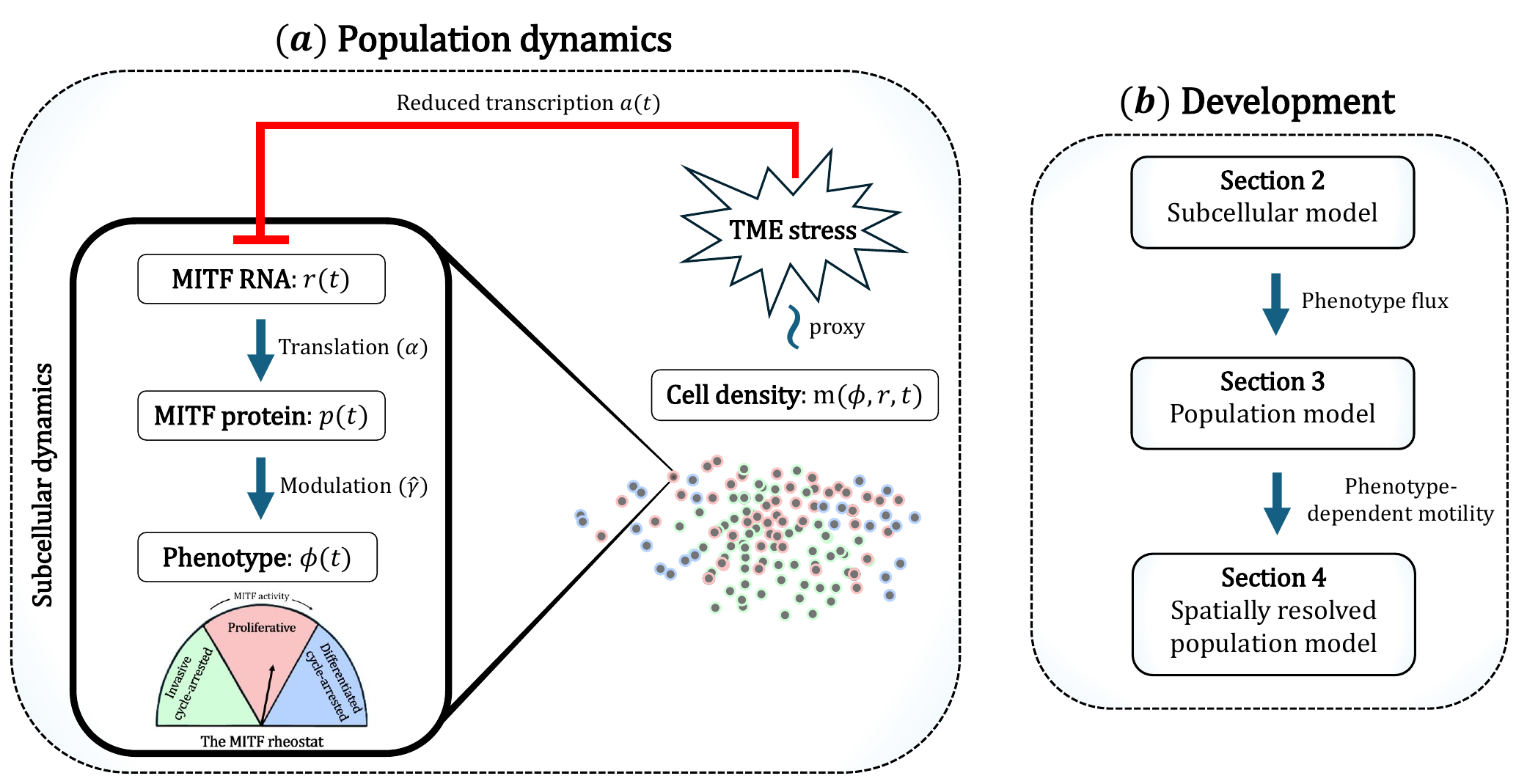}
    \caption{\textbf{A schematic of the multiscale, phenotype-structured population model and its incremental development. } (a) The model couples the subcellular dynamics for MITF RNA, protein and a phenotype variable to the time-evolution of a cell population density via a phenotype-dependent proliferation rate consistent with the MITF rheostat. The total density is used as an inverse proxy for the degree of microenvironment stress and informs the transcription of subcellular RNA. (b) The model development proceeds as shown. }
    \label{fig: schematic}
\end{figure}

We investigate the key question stated above using mathematical modelling. 
The model is formulated as a multiscale and phenotype-structured PDE that we develop and analyse incrementally.
In Sect. \ref{sec: subcellular}, we first consider subcellular dynamics for MITF RNA, protein and a downstream phenotype variable. 
By exploiting a separation of timescales, we reduce the subcellular model to a homogenised phenotype flux.
This flux informs the phenotype dynamics of a spatially homogeneous but phenotype-structured population model in Sect. \ref{sec: population}.
Finally, we consider a radially-resolved extension to the population model in Sect. \ref{sec: spatial}. 
The model is deliberately kept as simple as possible throughout so that we may calibrate parameters to published data via Bayesian inference.
A schematic of the model and its development is provided in Fig. \ref{fig: schematic}.

\section{Subcellular MITF dynamics} \label{sec: subcellular}








We first consider the dynamics of MITF within individual melanoma cells. The model is formulated as a stochastic system of differential equations which couples the following key quantities: \\
\begin{itemize}
    \item $r(t) =$ MITF RNA concentration;
    \item $p(t) =$ MITF protein concentration;
    \item $\phi(t) =$ MITF activity ($\sim$ phenotype). \\
\end{itemize}
The dynamics of $r(t)$ and $p(t)$ are calibrated to data from single-cell RNA sequencing and half-life measurements. We define the phenotype variable $\phi(t)$ as a quantity downstream of $p(t)$ that evolves on a slower timescale than $r(t)$ and $p(t)$. This separation of timescales allows us to reduce the full system to an effective phenotype flux that we incorporate into the structured population models of Sections \ref{sec: population} and \ref{sec: spatial}. 

\subsection{RNA and protein} \label{sec: RNA and protein}

We propose that the RNA and protein dynamics satisfy the following equations:
\begin{align}
    dr &= [\underbrace{a}_{\text{transcription}}  - \underbrace{\lambda_r r}_{\text{degradation}}] dt + \underbrace{\sqrt{2 \theta r^{q}} dW}_{\text{empirical noise}}, \label{eqn: r}\\
    dp &=  [\, \, \underbrace{\alpha r}_{\text{translation}}  \, \,- \underbrace{\lambda_p p}_{\text{degradation}}]dt. \label{eqn: p}
\end{align}

In Eq. \eqref{eqn: r} we assume that RNA is transcribed at an average rate $a$ and degrades at rate $\lambda_r$.
The final term accounts for stochasticity in RNA abundance that arises from the noisy subcellular environment.
Here $dW \sim \text{Normal}(0, dt)$ denotes an increment of a Wiener process.
The state-dependent diffusivity $\theta r^{q}$, where $\theta, q >0$, is an empirical form motivated by single-cell sequencing data. 

In Eq. \eqref{eqn: p} we assume that protein is produced at rate $\alpha$ per unit RNA and degrades at rate $\lambda_p$. 
For simplicity, we do not include an additional noise term for protein.

In what follows we allow the transcription rate to depend on time in a bounded manner such that $a_\text{max} := \sup_{t \geq 0} a(t) < \infty$. This introduces a natural scaling for $r$ and $p$:
\begin{align}
    &\hat{r} := \Big(\frac{a_\text{max}}{\lambda_r}\Big)^{-1} r, & &\hat{p} := \Big( \frac{\alpha \, a_\text{max}}{\lambda_r \lambda_p} \Big)^{-1} p.
\end{align}
We further define 
\begin{align}
    &\tau := \frac{t}{[t]}, & &[t] = \text{1 hour},
\end{align}
so that time is measured in units of hours and parameter values are approximately of order 1. This scaling transforms Eqs. \eqref{eqn: r} and \eqref{eqn: p} into:
\begin{align}
    d\hat{r} &= \hat{\lambda}_r (\hat{a} - \hat{r}) d\tau + \sqrt{2 \hat{\theta} \, \hat{r}^{q}}d\hat{W}, \label{eqn: rhat}\\
    d\hat{p} &= \hat{\lambda}_p (\hat{r} - \hat{p}) d\tau, \label{eqn: phat}
\end{align}
where 
\begin{align}
    &\hat{a} := \frac{a [t]}{a_\text{max}} \leq 1, & \hat{\theta} := \theta \lambda_r^{-1} a_\text{max}^{q-2}, & &\hat{\lambda}_r := \lambda_r [t], & &\hat{\lambda}_p = \lambda_p [t]
\end{align}
and $d\hat{W} \sim \text{Normal}(0, d\tau)$. For notational clarity, henceforth we omit the hats and assume all quantities are in dimensionless form. 

\subsubsection{Time-dependent solutions}
\begin{figure}[h]
    \centering
    \includegraphics[width=0.99\linewidth]{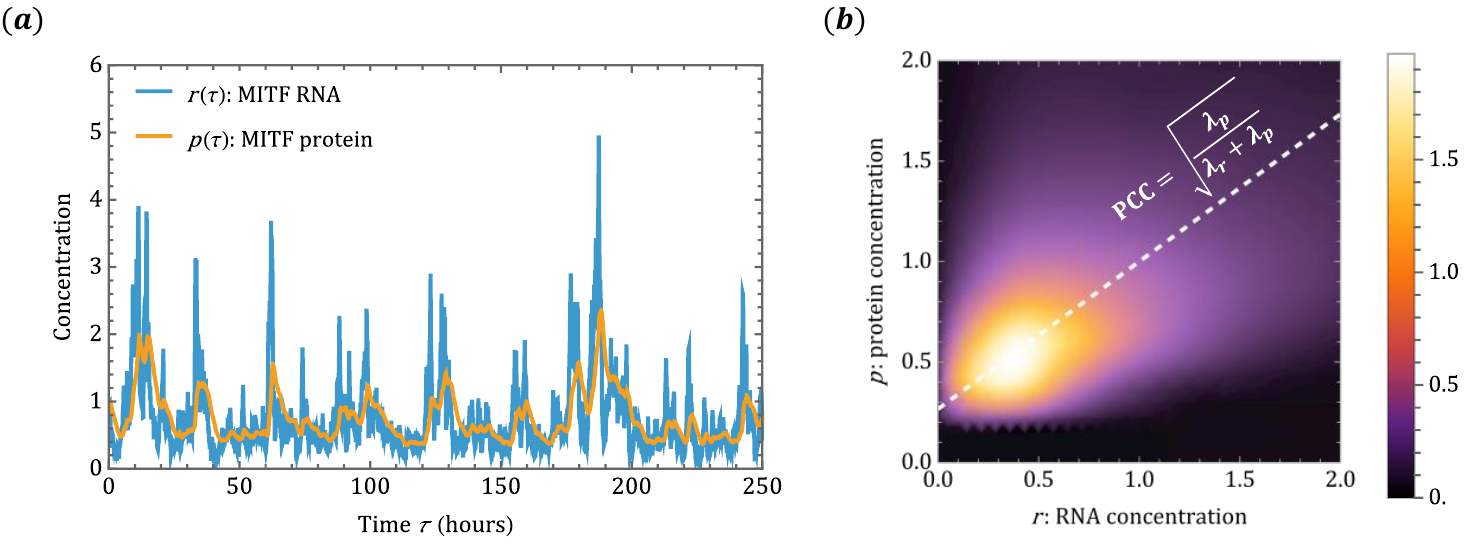}
    \caption{\textbf{Correlated and stochastic dynamics of MITF RNA $r(\tau)$ and protein $p(\tau)$ as defined by Eqs. \eqref{eqn: rhat} and \eqref{eqn: phat}. } Plot (a) shows an example time-dependent solution for $a = 1$. Plot (b) is a probability density map of the same solution $(r(\tau), p(\tau))$ for $0 < \tau < 10^5$. The overlaid dashed line passes through the mean $(1,1)$ with slope given by the Pearson Correlation Coefficient (PCC) specified in Eq. \eqref{eqn: PCC}. Parameter values are set to: $q =1.83$, $\theta = 0.20$, $\lambda_r = 0.28$, $\lambda_p = 0.35$.}
    \label{fig: RNA_protein_dynamics}
\end{figure}

Fig. \ref{fig: RNA_protein_dynamics} shows example time-dependent solutions to Eqs. \eqref{eqn: rhat} and \eqref{eqn: phat}. 
Here we set $a = 1$, which corresponds to the maximal RNA transcription rate that cells in our population models will exhibit in minimal stress conditions. 
The remaining parameters are set to values inferred in Sect. \ref{sec: subcellular_inference}. 

The trajectories $r(\tau)$ and $p(\tau)$ are overlaid in Fig. \ref{fig: RNA_protein_dynamics}(a). Both variables fluctuate asymmetrically about the mean ($a = 1$) with stochastic bursts. 
This behaviour is driven by the form of the RNA noise, which increases with $r$; cells with higher values of $r$ are more likely to take larger jumps that may drive them towards even higher values of $r$. The value of $p$ follows that of $r$ with a smoother trajectory since its stochasticity is purely inherited from $r$. 
Fluctuations in $p$ are visible on a timescale of hours while those of $r$ occur on a minute timescale. 
This smoother trajectory gives rise to less variation in $p$ overall.

Fig. \ref{fig: RNA_protein_dynamics}(b) shows a density plot in $(r,p)$ space of the same solution. The plot makes clear the positive correlation between $r$ and $p$ that emerges over time. 
Indeed, the distribution is highly symmetric about the overlaid line which passes through the mean $(1,1)$ with slope given by the Pearson Correlation Coefficient. 
The plot further highlights the skewness in the distribution; the mode occurs at values $r,p <1$ and there is a large tail. 

\subsubsection{Equilibrium distribution} \label{sec: rp equilibrium}

We are particularly interested in subcellular MITF dynamics when the average transcription rate $a = a(t)$ varies on the longer timescale of population dynamics. We formalise this timescale separation in Sect. \ref{sec: phenotype} and in our coupling of $a(t)$ to the cell density in the population models of Sects. \ref{sec: population} and \ref{sec: spatial}. The leading-order RNA and protein dynamics in this limit are given by setting $a = $ constant. We therefore pause to gain intuition by analysing the dynamics of Eqs. \eqref{eqn: rhat} and \eqref{eqn: phat} under this assumption. 

Since Eqs. \eqref{eqn: rhat} and \eqref{eqn: phat} define a stochastic process, we may recast the dynamics in terms of the Fokker-Planck equation, which, at equilibrium, reads
\begin{align}
    \frac{\partial}{\partial r}\big[ \lambda_r  (a-r)\pi - \partial_r (\theta r^{q} \pi)  \big] +   \frac{\partial}{\partial p} \big[\lambda_p(r-p) \pi\big] = 0. \label{eqn: pi}
\end{align}
Here $\pi(r,p)\geq0$ is the probability density for the cell to be in the state with RNA concentration $r\geq0$ and protein concentration $p\geq0$. We note that $\pi$ satisfies no-flux boundary conditions 
\begin{align}
    &\bigg[\lambda_r(a-r)\pi - \frac{\partial}{\partial r} (\theta r^q \pi)\bigg]_{r \rightarrow0,\infty} = 0, & &\pi_{p=0} = 0,
\end{align}
to conserve the normalisation $\int_0^\infty \int_0^\infty \pi drdp = 1$. Although an explicit solution for $\pi$ in terms of elementary functions is not available, it is straightforward to compute the moments:
\begin{align}
    &\langle r \rangle =  \langle p \rangle _\pi = a, & &\langle p^2 \rangle =  \langle rp \rangle = \frac{\lambda_r a^2 + \lambda_p \langle r^2 \rangle}{\lambda_r + \lambda_p}, \label{eqn: pi moments}
\end{align}
by integrating Eq. \eqref{eqn: pi} after multiplying by $1$, $rp$ and $p^2$. Moreover, using that the marginal distribution with respect to $r$ decouples from $p$, we may solve for it explicitly and calculate
\begin{align}
    \langle r^2 \rangle = \frac{\int_0^\infty r^{2-q} \exp \big[ \frac{\lambda_r}{\theta} r^{1-q} \big( \frac{a}{1-q} - \frac{r}{2-q} \big) \big] dr}{\int_0^\infty r^{-q} \exp \big[ \frac{\lambda_r}{\theta} r^{1-q} \big( \frac{a}{1-q} - \frac{r}{2-q} \big) \big] dr}, \qquad q \neq 1, 2. \label{eqn: <r^2>}
\end{align}
The algebraic singularities at $q = 1, 2$ in Eq. \eqref{eqn: <r^2>} are removable and the limiting expressions are obtained by continuity. Hence, $\langle r^2 \rangle$ is continuous for $q>0$. Combining these results, we obtain explicit expressions for the variances
\begin{align}
    &\text{Var}_r(a) = \langle r^2 \rangle - a^2, & &\text{Var}_p(a) = \frac{\lambda_p}{\lambda_p + \lambda_r} \text{Var}_r(a), \label{eqn: variances}
\end{align}
and the Pearson Correlation Coefficient (PCC)
\begin{align}
    \text{PCC} = \sqrt{\frac{\lambda_p}{\lambda_p + \lambda_r }} > 0. \label{eqn: PCC}
\end{align}
Eqs. \eqref{eqn: variances} and \eqref{eqn: PCC} make clear that the variance of $p$ relative to that of $r$ and the positive correlation between $r$ and $p$ are both determined by the relative decay rates in the combination $\lambda_p/(\lambda_p + \lambda_r)$. In particular, $\text{Var}_p(a) < \text{Var}_r(a)$ for all $a$.

\begin{figure}
    \centering
    \includegraphics[width=0.99\textwidth]{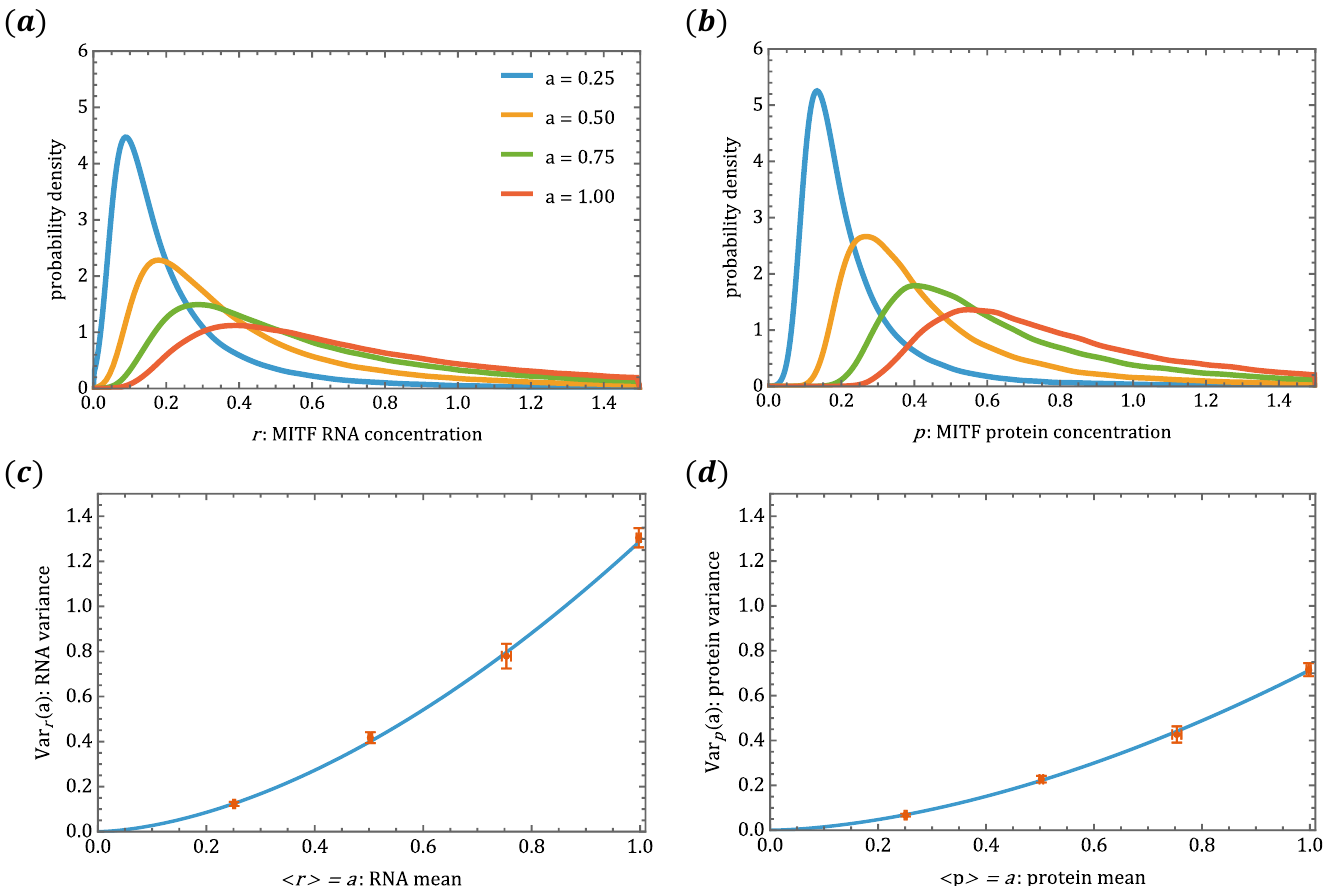}
    \caption{\textbf{Increases to the transcription rate, $a$, give rise to higher variances in the RNA and protein equilibrium distributions. } Plots (a) and (b) show kernel density estimates of the steady state RNA and protein concentrations, respectively. Each distribution is derived from a simulation of Eqs. \eqref{eqn: rhat} and \eqref{eqn: phat} for $0< \tau < 5 \times 10^4$. Plots (c) and (d) show the analytical expression for the mean-variance relationships of RNA and protein, respectively, with overlaid data from stochastic realisations at $a = 0.25$, $0.50$, $0.75$ and $1.00$. The points and uncertainties correspond to the mean $\pm$ SEM across $10$ simulations for $0<\tau < 2 \times 10^4$ at each value of $a$.}
    \label{fig: rna protein variance}
\end{figure}
We validate Eqs. \eqref{eqn: variances} and \eqref{eqn: PCC} by comparing to densities derived from numerical solutions to Eqs. \eqref{eqn: rhat} and \eqref{eqn: phat} for the fixed values $a = 0.25$, $0.50$, $0.75$, $1.00$. Fig. \ref{fig: rna protein variance} illustrates the comparison. Plots (a) and (b) show the marginal distributions for $r$ and $p$, respectively, which are kernel density estimates of simulations for $0 < \tau < 5 \times 10^4$ sampled at intervals $\Delta \tau = 0.1$ (i.e., every $6$ minutes). As expected, the distributions are increasingly concentrated towards zero as $a$ decreases. The lower variance of $p$ relative to $r$ is also visible in the larger values of the maxima for $p$ relative to the corresponding maxima in $r$. Plots (c) and (d) directly overlay the analytical predictions for $\text{Var}_r(a)$ and $\text{Var}_p(a)$ in Eqs. \eqref{eqn: variances} with the numerical estimates for the means and variances. The numerical results are consistent with the analytical predictions and show that the variance grows with the mean in a convex manner.   

\subsubsection{Parameter calibration} \label{sec: subcellular_inference}

Eqs. \eqref{eqn: rhat} and \eqref{eqn: phat} may serve as a generic model for a number of genes. 
To ensure specificity to MITF, we use published data to calibrate parameter values using a Bayesian likelihood-based approach.
The inference problem incorporates three independent datasets that we outline below. 
\begin{itemize}
    \item \textbf{Single-cell RNA sequencing (GSE72056, \cite{tirosh2016dissecting}). }We computed the (mean, variance) pairs of the MITF counts from the sequencing data of patients with untreated melanoma. 
For consistency with our model, we uniformly and linearly scaled the raw counts such that the patient with highest counts has unit mean. 
We label these observed data as $(a_i, V_i)$. The likelihood of each observation pertains to a lognormal observation model of the form
\begin{align}
    &\ln V_i = \ln\text{Var}_r(a_i) + \eta_1, & &\eta_1 \sim \text{Normal}(0, \sigma_1^2). \label{eqn: scseq_model}
\end{align}
\item \textbf{RNA decay \cite{goswami2015microrna}. }We re-digitised the time-course decay data of the wild-type mice experiments. In this study the authors stopped transcription using doxycycline and measured the proportion of remaining MITF RNA after four and eight hours. We label the data as $(\tau_i, r_i)$ and fit to exponential decay with lognormal observation noise, such that
\begin{align}
    &\ln r_i = \ln (e^{-\lambda_r \tau_i}) + \eta_2, & &\eta_2 \sim \text{Normal}(0, \sigma_2^2). \label{eqn: rna_decay_model}
\end{align}
\item \textbf{Protein half-life \cite{vu2024novel}. } We re-digitised the half-life measurements for MITF protein in the wild-type mice experiments. A complexity is that the authors observed higher stability in MITF that has been phosphorylated at amino acid location 73 (ps73) compared to proteins where that site is unphosphorylated (ups73). We label the data as $T_{ps73,i}$ and $T_{ups73, j}$ and fit to separate exponential decay rates with lognormal observation noise:
\begin{align}
    &\begin{cases}
        \ln T_{\text{ps73},i} = \ln ( 
        \ln 2/ \lambda_{\text{ps73}}) + \eta_3, \\
        \ln T_{\text{ups73},j} = \ln ( \ln 2/ \lambda_{\text{ups73}}) + \eta_3,
    \end{cases}
     & &\eta_3 \sim \text{Normal}(0, \sigma_3^2). \label{eqn: protein_decay_model}
\end{align} 
\end{itemize}
To construct an aggregate log-likelihood, we assume all data above are independent and sum the log-likelihoods of the individual observations according to Eqs. \eqref{eqn: scseq_model}-\eqref{eqn: protein_decay_model}. We supply all parameters with a uniform and positive prior distribution. The log-posterior consequently differs from the log-likelihood only by a constant. 

\begin{figure}
                \centering \includegraphics[width=0.99\textwidth]{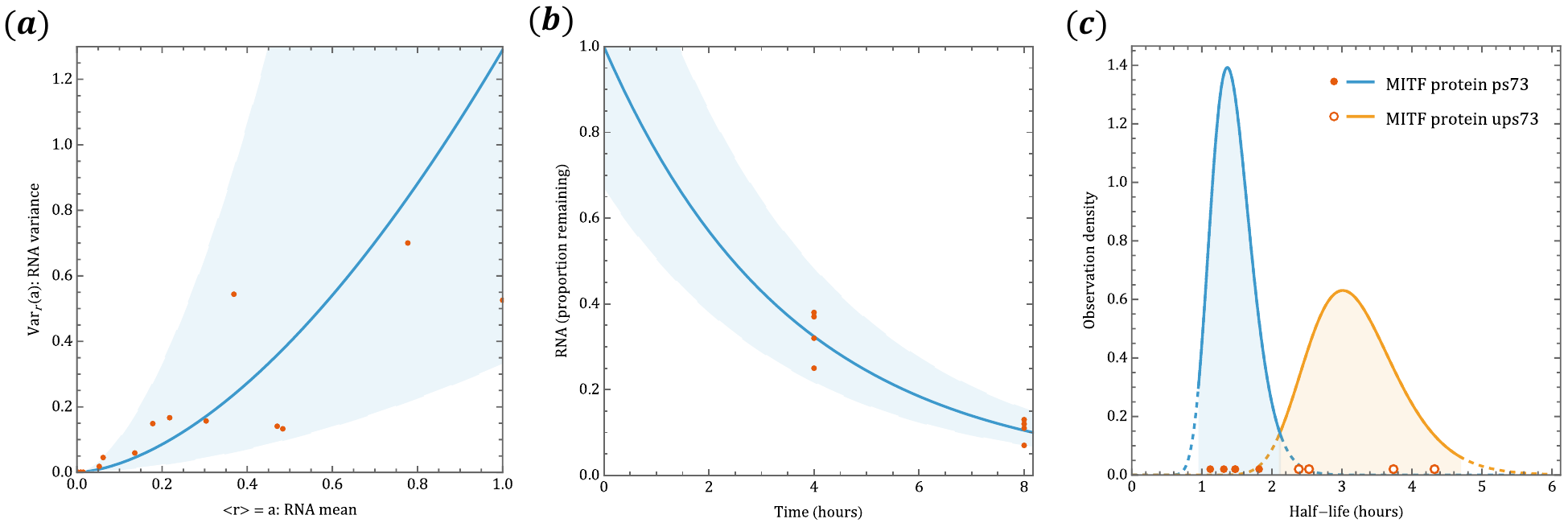}
                \caption{\textbf{Parameter values of the subcellular MITF model are calibrated to published data. The best fit (MAP) solution shows good agreement. } Plot (a) shows $\text{Var}_r(a)$ with overlaid data of normalised MITF counts from single-cell RNA sequencing \cite{tirosh2016dissecting}. Plot (b) shows the fitted exponential decay of RNA with overlaid data from \cite{goswami2015microrna}. Plot (c) shows the fitted lognormal observation distribution for MITF protein half-life in the cases where amino acid location 73 is phosphorylated (ps73) and unphosphorylated (ups73) with overlaid data from \cite{vu2024novel}. The shaded regions in all plots indicate where $95\%$ of observations are predicted to lie. The parameter values correspond to the MAP values listed in Eq. \eqref{eqn: subcellular_MAP}.}
                \label{fig: subcellular_MAP}
            \end{figure}

We first locate the parameter values that provide the best fit to the data in the sense that the posterior distribution is maximised. That is, we compute the \textit{maximum a posteriori} (MAP) estimate. An application of the Mathematica \textit{NMaximize} routine to the log-posterior yields 
\begin{align}
    \begin{split}
        &q = 1.83, \qquad \theta = 0.20, \qquad \lambda_r = 0.28, \qquad \lambda_{\text{ps73}} = 0.22, \\
        &\sigma_1 = 0.70, \quad \, \, \, \sigma_2 = 0.21, \quad \, \, \, \sigma_3 = 0.21, \qquad \lambda_{\text{ups73}} = 0.49. 
    \end{split} \label{eqn: subcellular_MAP}
\end{align}
Fig. \ref{fig: subcellular_MAP} shows a comparison between the MAP predictions and the experimental data.
The shaded regions in each plot indicate where $95\%$ of observations are predicted to lie according to Eqs. \eqref{eqn: scseq_model}-\eqref{eqn: protein_decay_model} when the parameter values are set to the values in Eq. \eqref{eqn: subcellular_MAP}. 
Fig. \ref{fig: subcellular_MAP}(a) shows that the function $\text{Var}_r(a)$ successfully captures the nonlinear RNA mean-variance relationship of the sequencing data when the RNA noise term in Eq. \eqref{eqn: rhat} grows with a sub-quadratic dependence on $r$. Likewise, we find predictions consistent with the data for RNA decay in Fig. \ref{fig: subcellular_MAP}(b) and protein half-life in Fig. \ref{fig: subcellular_MAP}(c). All data lie within the predicted $95\%$ observation regions.

\begin{figure}
    \centering
    \includegraphics[width=0.99\textwidth]{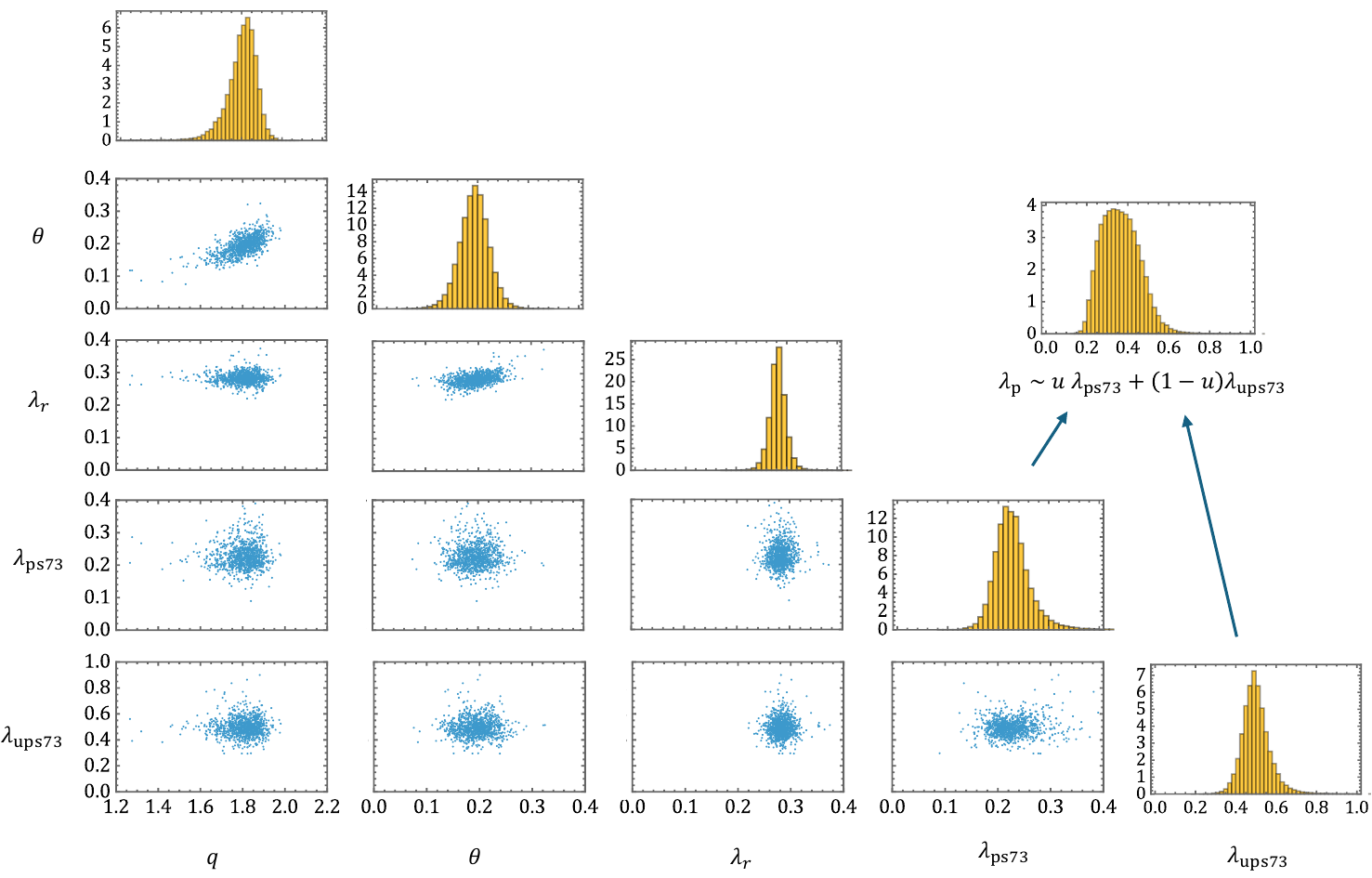}
    \caption{\textbf{Posterior distribution samples quantify uncertainty in the subcellular model parameter values. } Histograms on the diagonal represent the marginal distribution of each parameter. Pairwise scatter plots are also provided to analyse correlations. Each parameter is practically identifiable and we note a positive correlation $\text{PCC}=0.61$ between $q$ and $\theta$. Also shown are derived samples for $\lambda_p$ when the proportion of ps73 and ups73 MITF within cells is unknown and uniformly distributed as $u \sim \text{Uniform}(0,1)$. Samples represent 4 independent chains with $5\times 10^4$ iterations. Pairwise plots with the observation noise parameters $\sigma_i$, $i=1$, 2, 3 are also given in Appendix \ref{Appendix: subcellular_MCMC}. }
    \label{fig: subcellular_MCMC}
\end{figure}
To quantify parameter value uncertainties, we implement Markov Chain Monte Carlo (MCMC) sampling of the posterior distribution using a Metropolis-Hastings algorithm in Mathematica. 
We use 4 independent chains with $5\times 10^4$ iterations each (including $10^4$ burn-in samples). 
The Gelman-Rubin statistic is less than $1.01$ for each parameter and chain trace plots appear static and mix well so that we are confident of convergence. 
Diagnostic plots including posterior predictions are given in Appendix \ref{Appendix: subcellular_MCMC}.  

Fig. \ref{fig: subcellular_MCMC} illustrates the posterior distribution samples for the key parameters.
The marginal distribution of each parameter is unimodal with relatively narrow tails, which suggests practical identifiability.
The only significant correlation amongst the parameters is between $q$ and $\theta$, with $\text{PCC} = 0.61$. 
This positive correlation is unsurprising since both parameters impact the magnitude of the RNA noise term in Eq. \eqref{eqn: rhat}; the reduction in RNA noise by increasing $q$ may be partially compensated for by increasing $\theta$.
Since the proportion $u \in [0,1]$ of ps73 and ups73 MITF protein in melanoma cells \textit{in vivo} is unknown, we estimate the uncertainty in the effective MITF protein degradation rate $\lambda_p$ by generating samples of the form $ \lambda_p \sim u \lambda_{\text{ps73}} + (1-u) \lambda_{\text{ups73}}$ for $u \in \text{Uniform}(0,1)$. 
The resulting histogram for $\lambda_p$ is also shown in Fig. \ref{fig: subcellular_MCMC}. 
Since the samples for $\lambda_\text{ps73}$ and $\lambda_\text{ups73}$ contain some overlap, we anticipate that non-uniform choices for the proportion $u$ would yield a qualitatively similar distribution for $\lambda_p$. Samples for the noise parameters, omitted in Fig. \ref{fig: subcellular_MCMC} for readability, also indicate practical identifiability and are shown with the complete pairwise plots in Fig. \ref{fig: subcellular_MCMC_full}. 

\subsection{Phenotype} \label{sec: phenotype}

We define the melanoma cell phenotype variable $\phi \geq 0$ via the ODE
\begin{align}
    \frac{d \phi}{d\tau} &=  \hat{\gamma} (p - \phi). \label{eqn: phenotype definition}
\end{align}
Eq. \eqref{eqn: phenotype definition} ensures that phenotype follows the dimensionless MITF protein concentration, $p(\tau)$, at a rate $\hat{\gamma} > 0$. We note that, despite the apparently deterministic definition, $\phi$ inherits the stochasticity from its coupling to $p$. 

Motivated by the observations of Hoek \textit{et al}. \cite{hoek2008vivo}, we assume that $\phi$ evolves on the slower ($\sim$ weeks/months) timescale of population dynamics rather than the ($\sim$ hours) timescale of subcellular RNA and protein concentrations. To formalise this separation of timescales, we define
\begin{align}
    &t = \epsilon \tau, & &\hat{\gamma} = \epsilon \gamma, & &\epsilon = \frac{[1 \text{hour}]}{[1 \text{month}]} = \frac{1}{720} \ll 1,
\end{align}
and recast the full subcellular SDE subsystem \eqref{eqn: rhat}, \eqref{eqn: phat}, \eqref{eqn: phenotype definition} into its equivalent Fokker-Planck formulation. If $\psi (r, p , \phi, t)$ denotes the probability density for the cell to be in the state $(r, p, \phi)$ at time $t$, we have
\begin{align}
    \epsilon \frac{\partial \psi}{\partial t} + (\mathcal{L}_0 + \epsilon \mathcal{L}_1) \psi = 0, \label{eqn: fokker-Planck psi}
\end{align}
where $\mathcal{L}_0$ and $\mathcal{L}_1$ are the differential operators
\begin{align}
    &\mathcal{L}_0 \psi := \frac{\partial}{\partial r} \bigg[ \lambda_r(a - r) \psi - \frac{\partial}{\partial r} (\theta r^q \psi) \bigg] + \frac{\partial}{\partial p} \bigg[ \lambda_p (r-p) \psi \bigg], \label{eqn: L0} \\
    &\mathcal{L}_1 \psi := \frac{\partial}{\partial \phi} \bigg[ \gamma (p - \phi) \bigg]. \label{eqn: L1}
\end{align}
We impose the following no-flux boundary conditions 
\begin{align}
    \bigg[ \lambda_r(a - r) \psi - \frac{\partial}{\partial r} (\theta r^q \psi) \bigg]_{r \rightarrow0, \infty} = \psi_{p = 0} = \psi_{\phi = 0} = 0. \label{eqn: psi bconds}
\end{align}

\subsubsection{Effective phenotype flux}

Using an asymptotic analysis of $\psi$ in the limit $\epsilon \rightarrow 0$, we show that the dynamics of $\phi$, as encoded by its marginal distribution 
\begin{align}
    \Psi(\phi ,t) := \int_0^\infty \int_0^\infty \psi (r,p, \phi, t) dr dp,
\end{align}
reduce to an effective local flux. The derivation is detailed in Appendix \ref{Appendix: phenotype flux}. Importantly, we assume throughout that the transcription rate varies smoothly on the slow timescale, such that
\begin{align}
    a = a(\epsilon \tau) = a(t).
\end{align}
This is consistent with its coupling to the cell density in the population models of Sects. \ref{sec: population} and \ref{sec: spatial}. 

Up to contributions of $\mathcal{O}(\epsilon)$, we find that $\Psi$ satisfies an advection-diffusion equation
\begin{align}
    \frac{\partial \Psi}{\partial t} + \frac{\partial}{\partial \phi} \big( v \Psi - \frac{\partial}{\partial \phi} (\hat{D} \Psi) \big) = 0, \label{eqn: phenotype effective}
\end{align}
where the effective velocity, $v$, and diffusivity, $\hat{D}$, satisfy
\begin{align}
    v(\phi, t) &= \gamma \bigg[ (a - \phi)  - \epsilon \, \frac{da}{dt} \bigg( \frac{1}{\lambda_r} + \frac{1}{\lambda_p} \bigg) \bigg], & \hat{D}(t) &=  \frac{\epsilon \gamma^2 \text{Var}_r(a)}{\lambda_r}. \label{eqn: v, D}
\end{align}
Here $\text{Var}_r(a)$ is the RNA variance defined by Eq. \eqref{eqn: variances}. 
We note that $\hat{D} = \mathcal{O}(\epsilon)$ and so Eq. \eqref{eqn: phenotype effective} is dominated by phenotype advection towards $\phi = a$ at rate increasing with $\gamma$. 
The $\mathcal{O}(\epsilon)$ velocity correction accounts for a lag in response to changes in $a(t)$; if $\frac{da}{dt} > 0$ then $r$ and $p$ are on average slightly less than their predicted quasi-equilibrium values for given $a(t)$, so that $\phi$ targets a value $p < a$ in accordance with Eq. \eqref{eqn: phenotype definition} (and vice-versa when $\frac{da}{dt} < 0$). 
The diffusivity is directly proportional to the variance of $r$ and quantifies how much RNA stochasticity propagates to the phenotype variable. 
We further note that $\hat{D}$ scales with $\gamma^2$, while $v$ scales as $\gamma$, and so $\gamma$ also influences the contribution of advection relative to diffusion.

\subsubsection{Equilibrium distribution}

\begin{figure}
    \centering
    \includegraphics[width=0.6\textwidth]{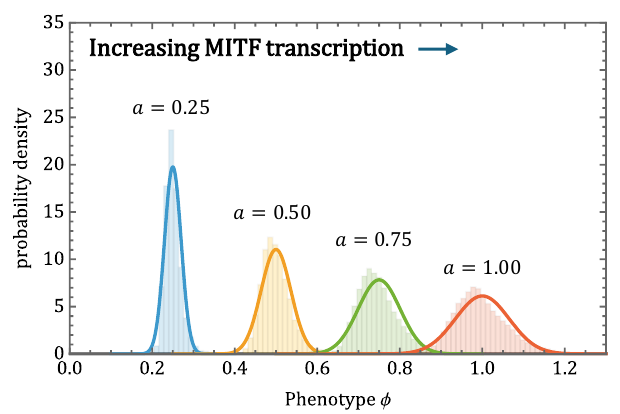}
    \caption{\textbf{Agreement between numerical and analytical phenotype distributions for individual cells at equilibrium. } We overlay the Gaussian solution Eq. \eqref{eqn: gaussian} on a histogram of time-dependent numerical solutions for $\phi(t)$ from the SDE system \eqref{eqn: rhat}, \eqref{eqn: phat} and \eqref{eqn: phenotype definition} for each $a = 0.25$, $0.5$, $0.75$, $1$. Note the increased variance in $\phi$ with $a$, which reflects the increased RNA variance in $a$ via Eq. \eqref{eqn: variances}. The histograms represent samples of $\phi(t)$ for each hour of a decade ($0 < t < 120$) with initial conditions $\phi(0) = a$. We set $\gamma = 0.67$ and fix the remaining parameter values at the MAP values in Eq. \eqref{eqn: subcellular_MAP} and $\lambda_p = 0.35$.}
    \label{fig: phenotype_gaussians}
\end{figure}

Eq. \eqref{eqn: phenotype effective} admits the exact equilibrium solution
\begin{align}
    &\Psi^\star(\phi) = C \cdot \exp \bigg[-\frac{\gamma (\phi - a)^2}{2\hat{D}}\bigg], & &C = \sqrt{\frac{2 \gamma}{\pi \hat{D}}}  \bigg[ 1 + \text{Erf}\big(a \sqrt{\frac{\gamma}{2\hat{D}}} \bigg) \bigg]^{-1}, \label{eqn: gaussian}
\end{align}
where the constant $C>0$ ensures that $\int_0^\infty \Psi^\star d \phi = 1$. Eq. \eqref{eqn: gaussian} describes the positive truncation of a  Gaussian with mean $\phi = a$ and standard deviation $\sqrt{\epsilon \gamma\text{Var}_r(a)/\lambda_r}$. 

In Fig. \ref{fig: phenotype_gaussians} we compare this analytical solution with histograms of time-dependent numerical solutions for $\phi(t)$ from the SDE system \eqref{eqn: rhat}, \eqref{eqn: phat} and \eqref{eqn: phenotype definition} for the cases $a = 0.25, 0.50, 0.75, 1.00$. Eq. \eqref{eqn: gaussian} agrees well with the numerical solutions for each value of $a$. The plots also highlight that solutions for higher values of $a$ vary more widely in $\phi$. This behaviour is captured by Eq. \eqref{eqn: gaussian} since $\text{Var}_r(a)$ is an increasing function and proportional to $\hat{D}$.



\section{Structured population model} \label{sec: population}

We have presented thus far a mathematical model for the time evolution of phenotype within individual melanoma cells. 
In reality, melanomas comprise a population of malignant cells that collectively influence patient outcome via their coupled dynamics.
We address this discrepancy in the current section by presenting a mathematical model for the time evolution of the phenotype distribution amongst melanoma cells within a tumour. 

The model is formulated as a partial differential equation (PDE) that is non-locally coupled with respect to phenotype.
The phenotype variable $\phi$ informs the proliferation rate in a manner consistent with the MITF rheostat. 
Importantly, by contrast to its usage as a given function or constant in Sect. \ref{sec: subcellular}, the average transcription rate $a(t)$ is coupled to the cell density as a proxy for the integrated stress response.
Spatial resolution is neglected here for simplicity  and explored in Sect. \ref{sec: spatial}.

\subsection{Model statement} \label{sec: structured population model statement}

Let $m(\phi, t) \geq 0$ be the density of melanoma cells with phenotype $\phi \geq 0$ at time $t \geq 0$. We propose that the dynamics are governed by the equation
\begin{align}
    \frac{\partial m}{\partial t} + \underbrace{\frac{\partial}{\partial \phi} \bigg[ v m - \frac{\partial}{\partial \phi} (\hat{D} m) \bigg]}_{\text{phenotype modulation}} &= \big[ \underbrace{\rho(\phi,t)}_{\text{proliferation}} - \underbrace{{\color{white}{\rho}}\nu {\color{white}{\rho}}}_{\text{death}} \big] m. \label{eqn: m dynamics}
\end{align}
The term $\partial_\phi [ vm - \partial_\phi (\hat{D}m)]$ accounts for changes in the population phenotype density due to phenotype modulation within individual melanoma cells. The coefficients of the flux, $v = v(\phi,t)$ and $\hat{D} = \hat{D}(t)$, are given by Eqs. \eqref{eqn: v, D} so that individual cell phenotype dynamics are consistent with the subcellular model developed in Sect. \ref{sec: subcellular}. It remains to specify how the dimensionless transcription rate $a(t) \leq 1$ is coupled to the population dynamics. We posit that 
\begin{align}
    &a(t) = \frac{1}{1 + M(t)/M^\star}, & &M(t) = \int_0^\infty m(\phi,t) d\phi, \label{eqn: a(t)}
\end{align}
where $M(t)$ is the total population density and $M^\star > 0$ is a constant. Eq. \eqref{eqn: a(t)} encodes the idea that MITF transcription rates decrease in response to cell stress and that cell density is a proxy for degree of stress. In particular, increases to cell density are likely to promote nutrient competition and mechanical confinement, which both drive MITF-low phenotype switching \cite{falletta2017translation, hunter2025mechanical}. 

The right-hand side terms in Eq. \eqref{eqn: m dynamics} account for the proliferation and death of melanoma cells. We assume that cells divide at a rate
\begin{align}
    \rho(\phi, t) = \rho_\text{max} \cdot \bigg( \frac{1}{1 + \kappa M(t) \, / M^\star} \bigg) \cdot W(\phi; \phi_L, \phi_R). \label{eqn: rho(phi)}
\end{align}
Here $W$ is a window function
\begin{align}
    W (\phi; \phi_L, \phi_R) = 
    \begin{cases}
        \sin^2 \Big[ \pi \Big(\frac{ \phi - \phi_L}{\phi_R - \phi_L} \Big) \Big], & \phi_L < \phi < \phi_R, \\
        0 & \text{otherwise}, \label{eqn: W}
    \end{cases}
\end{align}
that is nonzero only within the interval $\phi_L < \phi < \phi_R$, where $\phi_L$ and $\phi_R$ are constants, and attains its maximal value $1$ at $\phi = \frac{\phi_L + \phi_R}{2}$. The functional form \eqref{eqn: W} reflects the MITF rheostat \cite{carreira2006mitf, goding2014fishful} in that only cells with intermediate MITF activity are proliferative. However, the specific  formulation as a symmetric sinusoidal window is an arbitrary modelling choice used for simplicity. The factor $(1+\kappa M(t)/M^\star)^{-1}$ in Eq. \eqref{eqn: rho(phi)} accounts for external limitations to the proliferation rate due to contact inhibition and metabolite deficiency, for which the cell density $M(t)$ is again a proxy. Hence, cells in the model only proliferate at the maximum rate $\rho_\text{max}$ when in the ideal transcriptional state ($\phi = \frac{\phi_L + \phi_R}{2}$) and contact inhibition is minimal ($M \ll M^\star$). Daughter cells are assumed to have the same phenotype as their parent cells. We further assume that cells die at a constant rate $\nu$ for simplicity. \\

\noindent \textbf{Boundary conditions.} We impose the no-flux boundary conditions
\begin{align}
    \bigg[ v m - \frac{\partial}{\partial \phi} (\hat{D}m) \bigg]_{\phi \rightarrow 0, \infty} = 0 \label{eqn: m noflux}
\end{align}
so that cells may not leave the domain via unrealistic fluxes at $\phi = 0$ or as $\phi \rightarrow \infty$. \\

\noindent \textbf{Explicit rates of change.} By integrating Eq. \eqref{eqn: m dynamics} and using Eq. \eqref{eqn: m noflux} to eliminate boundary terms, we find the rate of change in total cell density
\begin{align}
    \frac{dM}{dt} = \int_0^\infty \big[ \rho(\phi, t) - \nu \big] m(\phi, t) d\phi
\end{align}
is the aggregate gain via proliferation less death. Hence, by applying the chain rule to Eq. \eqref{eqn: a(t)}, the derivative which appears in Eq. \eqref{eqn: v, D} may also be written explicitly as 
\begin{align}
    \frac{da}{dt} = - \frac{\frac{dM}{dt}}{M^\star (1+ M/M^\star)^2} = - \frac{\int_0^\infty \big[ \rho(\phi, t) - \nu \big] m(\phi, t) d\phi}{M^\star (1+ M/M^\star)^2}. \label{eqn: da/dt}
\end{align}

\noindent \textbf{Initial conditions.} We close the model with initial conditions
\begin{align}
    &m(\phi, 0) = M_0 \cdot \hat{m}_0 (\phi), & &M_0 \ll M^\star,& &\int_0^\infty \hat{m}_0(\phi) d\phi = 1. \label{eqn: m init}
\end{align}
The restriction $M_0 \ll M^\star$ ensures that $\vert 1 - a(0) \vert \ll 1$ by Eq. \eqref{eqn: a(t)}, and so the initial cell density is small enough that stress-induced reduction of MITF transcription is negligible. \\

\noindent \textbf{Non-dimensionalisation.} For numerical solutions we recast the model in terms of the dimensionless variables
\begin{align}
    &\tilde{m}(\phi, t) = \frac{m(\phi, t)}{M^\star}, & &\tilde{M}(t) = \frac{M(t)}{M^\star}. \label{eqn: m nondim}
\end{align}
The scaling \eqref{eqn: m nondim} measures melanoma cell densities in units of $M^\star$, the density for which MITF transcription is half-maximal. Dropping the tilde adornments, the dimensionless equations are identical to Eqs. \eqref{eqn: m dynamics}-\eqref{eqn: m init} under the replacement $M^\star \rightarrow 1$ and omitted for brevity.

\subsection{Time-dependent solutions} \label{sec: structured population dynamics}

We generate numerical solutions to Eqs. \eqref{eqn: m dynamics}-\eqref{eqn: m nondim} using the method of lines. Specifically, we use a uniform second-order central differencing in $\phi$ across a finite computational domain, typically $0 < \phi < 2$, with uniformly distributed nodes of separation $d\phi = 0.01$. The integral $M(t)$ is approximated using the trapezoidal rule and the resulting system of ODEs is solved via the \textit{NDSolve} routine in Wolfram Mathematica. 

\begin{figure}
    \centering
    \includegraphics[width=0.99\textwidth]{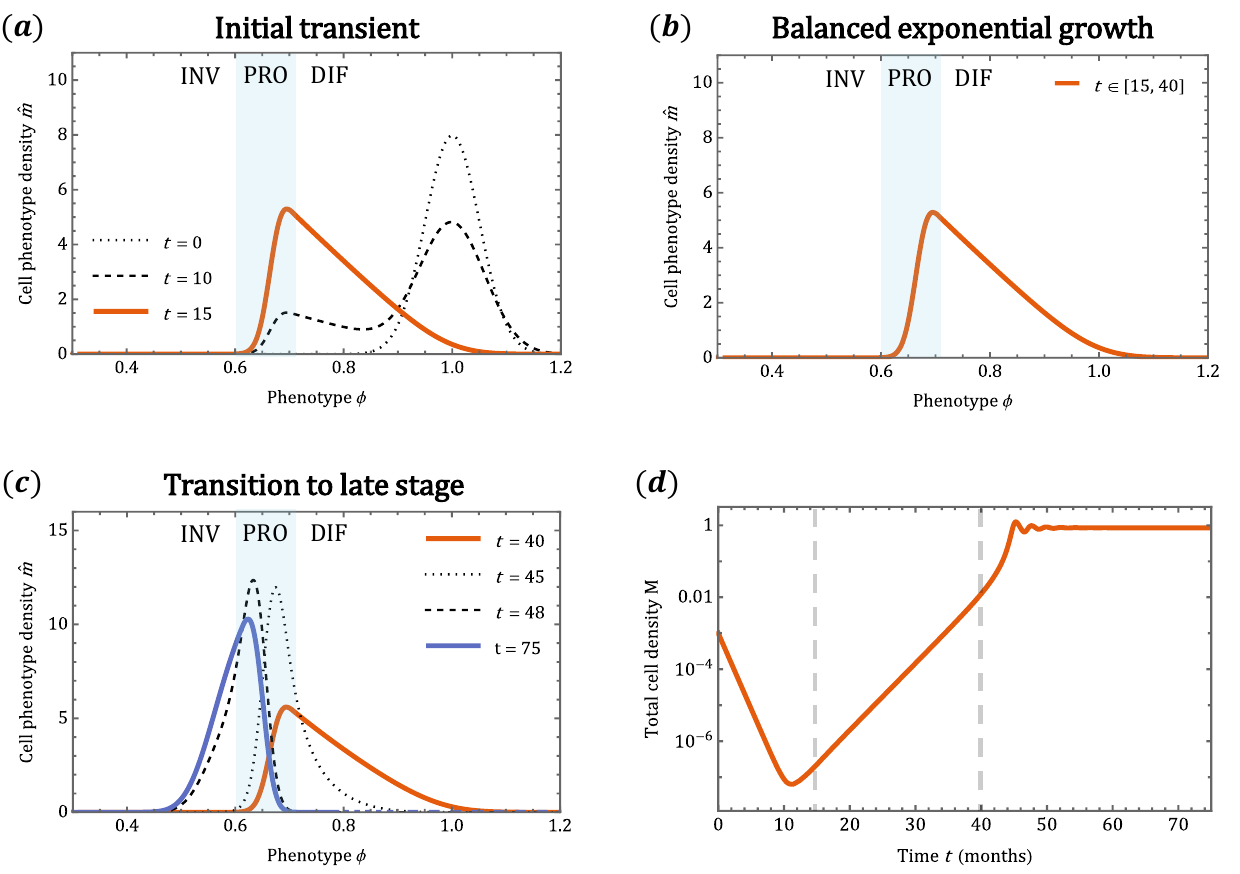}
    \caption{\textbf{Typical population dynamics for the model melanoma. }The dynamics defined by Eqs. \eqref{eqn: m dynamics}-\eqref{eqn: m nondim} may be considered in three stages: (a) an early transient that depends on the initial phenotype distribution, (b) a period of balanced exponential growth and (c) a transition towards late stage behaviour (e.g., steady state). The blue shading depicts the proliferative window $\phi \in (\phi_L, \phi_R)$. Plot (d) shows the time evolution of the total cell density, $M(t)$. Note the logarithmic vertical axis. Vertical lines separate the growth stages. The initial phenotype distribution is Gaussian and $M_0 = 10^{-3}$. Parameter values are set to: $\phi_L = 0.60$, $\phi_R = 0.71$, $\rho_\text{max} = 10.4$, $\nu = 0.97$, $\gamma = 0.67$, $q = 1.79$, $\theta = 0.19$, $\lambda_r = 0.28$, $\lambda_p = 0.35$ and $\kappa = 2$. }
    \label{fig: population_dynamics}
\end{figure}
Typical solutions of the model \eqref{eqn: m dynamics}-\eqref{eqn: m nondim} are shown in Fig. \ref{fig: population_dynamics}. 
Plots (a), (b) and (c) show the normalised cell phenotype density $\hat{m}(\phi, t) := m(\phi, t) / M(t)$ over time, while plot (d) illustrates the dynamics of the total cell density $M(t)$. The dynamics of the model melanoma may be considered in three stages.

At early times there is an initial transient that depends on the initial phenotype distribution (see Fig. \ref{fig: population_dynamics}(a)).
The initial condition rapidly washes out in favour of a phenotype distribution $\hat{m}_\text{exp}(\phi)$ that consists primarily of differentiated $\phi > \phi_R$ and proliferative $\phi_L < \phi < \phi_R$ cells. 
We verified numerically that the same phenotype profile $\hat{m}_\text{exp}(\phi)$ emerges independently of the initial phenotype distribution, $\hat{m}_0(\phi)$, provided that $M_0 \ll 1$ is sufficiently small.  
Examples for several Gaussian initial conditions are provided in Fig. \ref{fig: early dynamics}. In general, initial conditions concentrated away from the proliferative window yield an early decay in total cell density, while initial conditions containing many proliferative cells sustain growth from $t = 0$. 

The phenotype profile $\hat{m}_\text{exp}(\phi)$ is sustained throughout a period of balanced exponential growth (see Fig. \ref{fig: population_dynamics}(b)). We may understand this behaviour mathematically by noting that Eq. \eqref{eqn: m dynamics} admits a separable solution in the low density limit $M \ll 1$ where $a(t) \sim 1$ and $\frac{da}{dt} \sim 0$. Direct substitution shows that  
\begin{align}
    m(\phi, t) = M(t) \cdot \hat{m}_\text{exp}(\phi) \label{eqn: m separable}
\end{align}
satisfies Eq. \eqref{eqn: m dynamics} in this limit provided that
\begin{align}
    &\frac{dM}{dt} = S M, &  &S:= \int_0^\infty [\tilde{\rho}(\phi) - \nu] \hat
    m_\text{exp}(\phi) d\phi, \label{eqn: S}
\end{align}
and $\hat{m}_\text{exp}(\phi)$ solves the eigenvalue problem
\begin{align}
    &\frac{\partial}{\partial \phi} \bigg( \tilde{v} \hat{m}_\text{exp} - \frac{\partial}{\partial \phi} (\tilde{D} \hat
    m_\text{exp}) \bigg) = (\tilde{\rho} - \nu - S) \hat{m}_\text{exp}, & &\int_0^\infty \hat{m}_\text{exp}(\phi) d\phi = 1. \label{eqn: exp growth pde}
\end{align}
Here $\tilde{v}$, $\tilde{D}$ and $\tilde{\rho}$ are the phenotype velocity, phenotype diffusivity and proliferation rate in the low density limit $M \ll 1$:
\begin{align}
    &\tilde{v}(\phi) := \gamma(1-\phi), & &\tilde{D} := \frac{\epsilon \gamma^2 \text{Var}_r(1)}{\lambda_r}, & &\tilde{\rho}(\phi) := \rho_\text{max} \, W(\phi; \phi_L, \phi_R). \label{eqn: v,D,rho tildes}
\end{align}
Distributions $\hat{m}_\text{exp}(\phi)$, as defined above, agree well with the phenotype distribution observed in the exponential growth phase of numerical solutions. The growth at rate $S$ is disrupted once $M(t) = \mathcal{O}(1)$ and Eq. \eqref{eqn: m separable} loses its validity. In practice, this disruption is noticeable when $M(t) \approx 10^{-2}$. 

\begin{figure}
    \centering
    \includegraphics[width=0.99\textwidth]{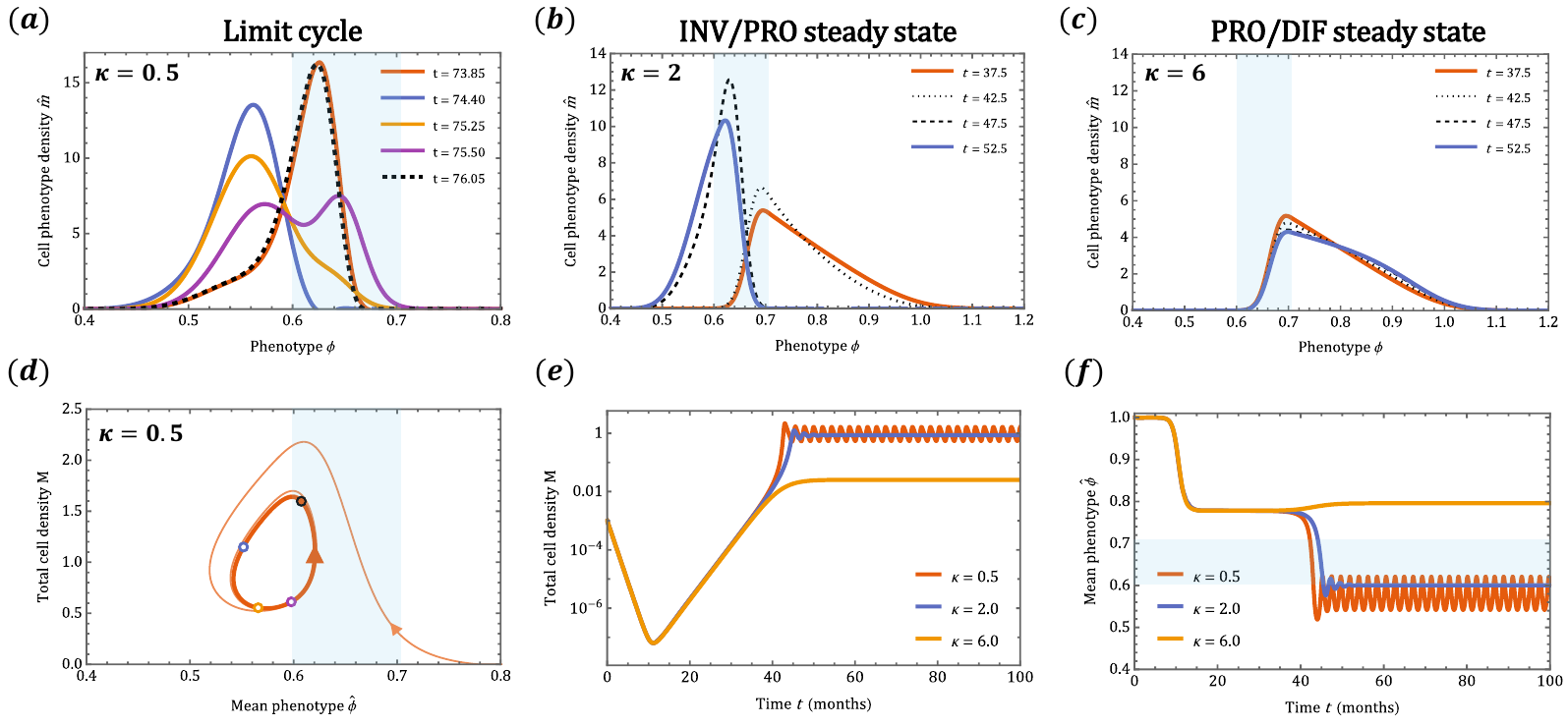}
    \caption{\textbf{Late-stage behaviours of the model melanoma. } Solutions to Eqs. \eqref{eqn: m dynamics}-\eqref{eqn: m nondim} exhibit three qualitatively distinct long-term behaviours: (a) a limit cycle marked by transitions between invasive and proliferative phenotypes, (b) a steady state with invasive and proliferative phenotypes and (c) a steady state with differentiated and proliferative cells. Plot (d) shows the limit cycle for $\kappa = 0.5$ in the $(\hat{\phi},M)$ plane. Plots (e) and (f) show the dynamics of total cell density $M(t)$ and mean phenotype $\hat{\phi}(t)$, respectively. Initial conditions and parameter values are identical to those in Fig. \ref{fig: population_dynamics} except for the value of $\kappa$, which is set as indicated.}
    \label{fig: 3longterm_behaviours}
\end{figure}
At later times when $M(t) = \mathcal{O}(1)$ the solution transitions towards a limiting behaviour (e.g., steady state in Fig. \ref{fig: population_dynamics}(c)). 
In general, the model exhibits three qualitatively distinct long-term behaviours that are illustrated in Fig. \ref{fig: 3longterm_behaviours}. 
First, the dynamics may tend towards a limit cycle (Fig. \ref{fig: 3longterm_behaviours}(a)) that periodically transitions between a proliferative-dominant and invasive-dominant population. 
This behaviour may be understood as follows. At the exit of the exponential growth phase, the population is sufficiently dense ($M = \mathcal{O}(1)$) to drive a stress-induced reduction in MITF transcription (recall Eq. \eqref{eqn: a(t)}) and thereby promote invasive phenotype switching. Once distanced from the proliferative window, the net rate of cell death exceeds that of proliferation, which causes the population to decay. This decay relieves the stress and consequently drives positive phenotype switching back into the proliferative window, starting the cycle anew.
Second, the dynamics can tend to a steady state balance between invasive and proliferative phenotypes (Fig. \ref{fig: 3longterm_behaviours}(b)).
In this case, the growth of cell density $M(t)$ at the exit of the exponential phase is sufficiently damped by contact inhibition to avoid phenotype switching that overshoots the proliferative window. 
Lastly, the dynamics may tend to a steady state balance between non-cycling, differentiated (DIF) cells and proliferative (PRO) cells (Fig. \ref{fig: 3longterm_behaviours}(c)).
In this case, contact inhibition is severe enough that $M(t)$ never rises sufficiently to drive invasive phenotype switching.
The long-term phenotype distribution is similar to $\hat{m}_\text{exp}(\phi)$ but exhibits a lower peak and larger tail. 
The explicit dynamics of the total cell density $M(t)$ and mean phenotype $\hat{\phi}(t) = \int_0^\infty \phi \hat{m}(\phi,t)d\phi$ are also given in Figs. \ref{fig: 3longterm_behaviours}(d)-(f). 
Fig. \ref{fig: 3longterm_behaviours}(d) makes clear that the limit cycle behaviour can be understood as anticlockwise rotation in $(\hat{\phi}, M)$ space with dynamics consistent with the explanation above. 
Figs. \ref{fig: 3longterm_behaviours}(e) and \ref{fig: 3longterm_behaviours}(f) highlight the contrast between the PRO/DIF steady state and the other long-term behaviours; the cell density is markedly lower and the mean phenotype increases upon exit of the exponential growth phase rather than becoming invasive. We confirm in Sect. \ref{sec: population long-term behaviours} that this behaviour reflects a branch separation with respect to the bifurcation parameter $\kappa$. 

\subsection{Parameter calibration} \label{sec: population model parameter calibration}

The model \eqref{eqn: m dynamics}-\eqref{eqn: m nondim} is formulated in terms of several parameters. 
To ensure the dynamics are representative of melanoma, we use published data to calibrate parameter values as a Bayesian inference problem. 
We use data from \textit{in vitro} and \textit{in vivo} experiments to construct informative prior distributions.
We compare clinical data to outputs of the model exponential growth phase, as defined in Eqs. \eqref{eqn: m separable}-\eqref{eqn: v,D,rho tildes}, using likelihood functions.
A brief outline of the data and their role in the inference problem is provided below.

The inference problem incorporates likelihoods derived from the following datasets.
\begin{itemize}
    \item \textbf{Ki-67 proliferation index \cite{ladstein2010ki}. }We re-digitised the measurements for the Ki-67 index derived from cutaneous melanoma patients. The Ki-67 index measures the percentage of sampled cells that are actively cycling. We label the data as $\mathcal{K}_i \in [0,1]$, and fit to a logit-normal observation model of the form 
    \begin{align}
        &\text{logit}(\mathcal{K}_i) = \text{logit}(\mathcal{K}_\text{model}) + \eta_4, & &\eta_4 \sim \text{Normal}(0, \sigma_4^2), \label{eqn: Ki-67 observations}
    \end{align}
    where
    \begin{align}
        &\text{logit}(x):= \ln \bigg( \frac{x}{1-x} \bigg), & &\mathcal{K}_\text{model} = \int_{\phi_L}^{\phi_R} \hat{m}_\text{exp}(\phi) d \phi. 
    \end{align}
    \item \textbf{Tumour doubling time \cite{eskelin2000tumor}. } We re-digitised the measurements for the doubling times derived from uveal melanoma patients. We label the data as $T_i$ and fit to a lognormal observation model of the form
    \begin{align}
        &\ln (T_i) = \ln (T_\text{model}) + \eta_5, & &\eta_5 \sim \text{Normal}(0, \sigma_5^2),
    \end{align}
    where
    \begin{align}
        T_\text{model} = \frac{\ln(2)}{S} = \frac{\ln(2)}{\int_0^\infty [\tilde{\rho}(\phi) - \nu] \hat{m}_\text{exp}(\phi) d\phi}. \label{eqn: T observations}
    \end{align}
\end{itemize}
We assume all data are independent and accordingly define the aggregate log-likelihood by summing the individual log-likelihoods of each data point. 

\begin{figure}
    \centering
    \includegraphics[width=0.99\textwidth]{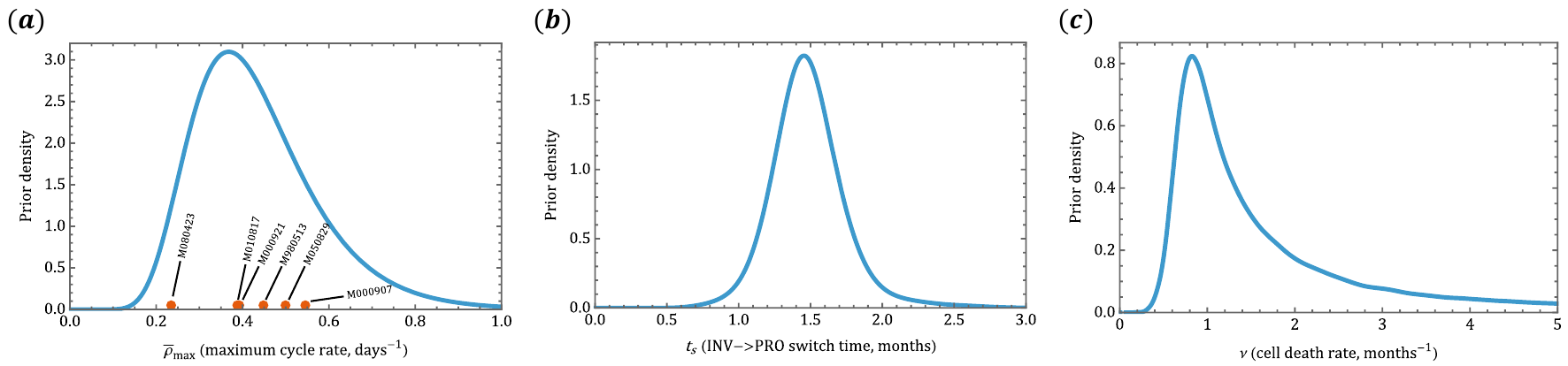}
    \caption{\textbf{Informative prior distributions for the population inference problem. } Plot (a) shows the lognormal prior for the maximum cycle rate $\tilde{\rho}_\text{max}$ with overlaid cell culture data. Plots (b) and (c) show the prior distributions for the phenotype switching time $t_s$ and cell death rate $\nu$, which are derived in Appendices \ref{Appendix: Hoek} and \ref{Appendix: TUNEL}, respectively. }
    \label{fig: informative_priors}
\end{figure}
The likelihood derived from the above data is insufficient to identify all parameter values when uniform positive prior distributions are assumed. We therefore incorporate further data to inform a prior distribution on parameter space. Specifically, we use
\begin{itemize}
    \item \textbf{Culture doubling times for proliferative cell lines \cite{widmer2012systematic}. }We use the doubling times for proliferative-phenotype melanoma cultures to construct a prior distribution on the maximum proliferation rate $\rho_\text{max}$. Specifically, we assume $\rho_\text{max} \sim \text{Lognormal}(\mu, \sigma^2)$ with $\mu = 2.51$ and $\sigma = 0.33$ so that  $\ln(2)/\rho_\text{max}$ is distributed with the same mean and variance as the culture data.
    \item \textbf{Xenograft invasive to proliferative switching times \cite{hoek2008vivo}. } Although direct measurements of phenotype switching rates \textit{in vivo} are not currently available, this study offers murine data for the lag in tumour initiation times when seeded with MITF-low versus proliferative human melanoma cells. 
    The authors attribute this lag to the time required for the MITF-low population to switch into a proliferative phenotype.
    Within the context of our mathematical model, we use this data to estimate the quantity
    \begin{align}
        t_s := -\gamma^{-1} \ln (1 - \phi_L). \label{eqn: ts}
    \end{align}
    Recalling that \eqref{eqn: m dynamics} is advection-dominant, $t_s$ is the time required for cells to drift from $\phi = 0$ into the proliferative window $\phi \in (\phi_L, \phi_R)$ by advection under minimal stress conditions. A derivation of the prior distribution for $t_s$ via Bayesian inference is supplied in Appendix \ref{Appendix: Hoek}. 
    \item \textbf{TUNEL index \cite{shukuwa2002fas}. }We re-digitised the TUNEL index data for cutaneous melanoma. The measurement represents the proportion of dead cells in the sample. Exploiting a separation of timescales and Bayesian inference, we use these measurements to derive an informative prior for the death rate, $\nu$, in  Appendix \ref{Appendix: TUNEL}. 
    \item \textbf{Non-invasive melanoma \textit{in situ} \cite{guerry1993lessons}. }We assume in the minimal stress conditions for which our model exhibits exponential growth that invasive cells do not constitute a significant proportion of the cell population (here taken as $<1\%$). Biologically, this is consistent with observations that many forms of melanoma (superficial spreading, lentigo maligna, acral lentiginous) initially undergo a non-metastatic radial growth phase. We enforce this during inference by rejecting proposed MCMC samples for which
    \begin{align}
        \int_0^{\phi_L} \hat{m}_\text{exp}(\phi) d \phi \geq 1\%.
    \end{align}
\end{itemize}

We cast the inference problem in terms of transformed variables to ensure parameter values are $\mathcal{O}(1)$ and to avoid unnecessary correlations. Specifically, we replace $(\phi_R, \rho_\text{max}, \gamma)$ with the variables $(\Delta \phi, \tilde{\rho}_\text{max}, t_s)$ where
\begin{align}
    &\Delta\phi := \phi_R - \phi_L, & &\tilde{\rho}_\text{max} := \frac{\rho_\text{max}}{30}. \label{eqn: inference variables}
\end{align}
Here $\Delta \phi$ is the width of the proliferative window,  $\tilde{\rho}_\text{max}$ is the maximum proliferation rate in days$^{-1}$ and $t_s$ the switching time defined in Eq. \eqref{eqn: ts}. The informative prior distributions for $\tilde{\rho}_\text{max}$, $t_s$ and $\nu$ are plotted in Fig. \ref{fig: informative_priors}. We assign uniform priors for $0 < \phi_L < 1$, $\Delta \phi > 0$ and the observation noises $\sigma_4, \sigma_5 >0$. The prior distribution of the subcellular parameters $(q, \theta, \lambda_r)$ are kernel density estimates of the subcellular posterior distribution (recall Fig. \ref{fig: subcellular_MCMC}); we import a 2-dimensional estimate for $(q, \theta)$ to capture their positive correlation and a one-dimensional estimate for $\lambda_r$. 

\begin{figure}
    \centering
    \includegraphics[width=0.99\textwidth]{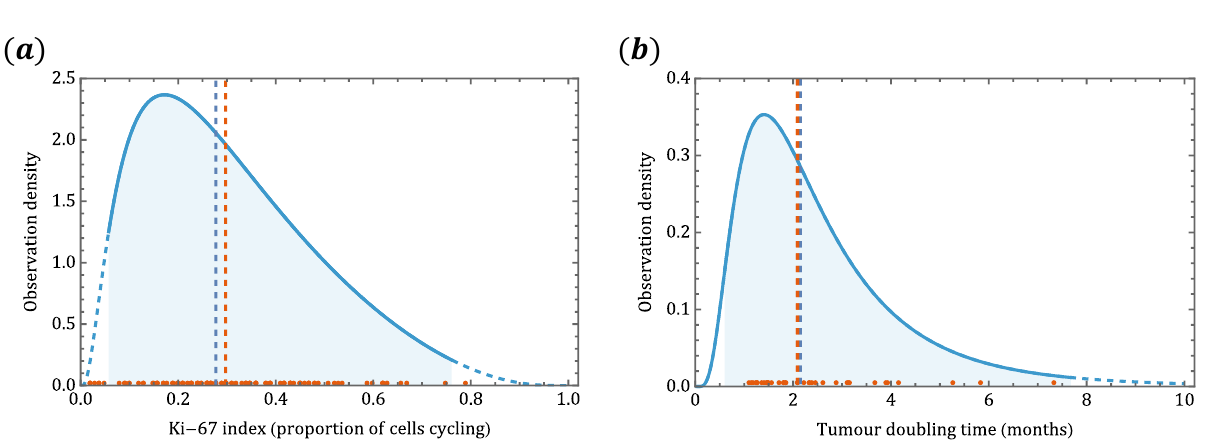}
    \caption{\textbf{The best fit (MAP) parameter values of the population model produce good agreement with published data. } Plot (a) shows the logit-normal observation distribution of the Ki-67 measurement, which represents the proportion of cells cycling, overlaid with the data from \cite{ladstein2010ki}. Plot (b) shows the lognormal observation distribution for the tumour doubling time  with overlaid data from \cite{eskelin2000tumor}. The shaded regions indicate where $95\%$ of observations are predicted to lie. The vertical lines correspond to the predicted (blue) and data (red) medians. 
    The parameter values correspond to the MAP values listed in Eq. \eqref{eqn: population_MAP}.}
    \label{fig: population_MAP}
\end{figure}
We first locate the MAP parameter values that provide the best fit  to the data. 
Applying the Mathematica \textit{NMaximize} routine to the log-posterior yields 
\begin{align}
    \begin{split}
        &\phi_L = 0.603, \quad \Delta \phi = 0.110, \quad \tilde{\rho}_\text{max} = 0.347, \quad \nu = 0.966, \quad t_s = 1.385, \\
        &\sigma_4 = 1.008, \quad \, \, \sigma_5 = 0.429, \quad \, \, q = 1.791, \quad \quad \, \, \theta = 0.186, \quad \lambda_r = 0.284.
    \end{split} \label{eqn: population_MAP}
\end{align}
Fig. \ref{fig: population_MAP} shows a comparison between the MAP predictions and the clinical data. 
The shaded regions in each plot indicate where $95\%$ of observations are predicted to lie according to Eqs. \eqref{eqn: Ki-67 observations}-\eqref{eqn: T observations} when the parameter values are set to the values in Eq. \eqref{eqn: population_MAP}. 
The plots indicate consistency between the observation distributions and overlaid data for Ki-67 (Fig. \ref{fig: population_MAP}(a)) and tumour doubling time (Fig. \ref{fig: population_MAP}(b)). 

To quantify parameter value uncertainties, we again implement Markov Chain Monte Carlo (MCMC) sampling of the posterior distribution using a Metropolis-Hastings algorithm in Mathematica. 
We use 6 independent chains with $2\times 10^6$ iterations each. 
The Gelman-Rubin statistic is less than $1.01$ for each parameter and chain trace plots appear static and mix well so that we are confident of convergence. 
Diagnostic plots including posterior predictions are given in Appendix \ref{Appendix: population_MCMC}.  

\begin{figure}
    \centering
    \includegraphics[width=0.99\textwidth]{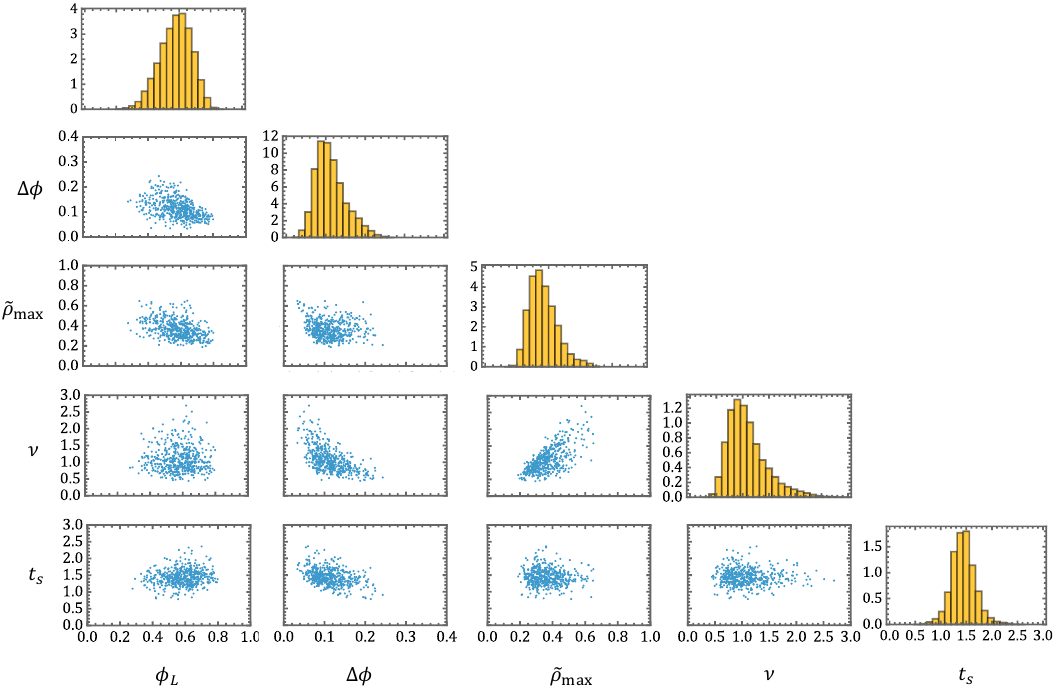}
    \caption{\textbf{Posterior distribution samples for the key parameters of the population model. }Histograms on the diagonal represent the marginal distribution of each parameter. Pairwise scatter plots are also provided to visualise correlations. We note that samples for all parameters are localised and yield bell-curve histograms. }
    \label{fig: population_MCMC}
\end{figure}
Fig. \ref{fig: population_MCMC} illustrates the posterior distribution samples for the key parameters.
The marginal distribution histograms for each parameter are localised and decay rapidly, which suggests practical identifiability. 
The modal values are consistent with the MAP estimate \eqref{eqn: population_MAP}. 
The pairwise plots reveal at least two significant correlations. 
First, the positive correlation between $\tilde{\rho}_\text{max}$ and $\nu$ is unsurprising since the net growth rate relies on the difference between the proliferation and death rates. 
There is also a negative correlation between $\Delta \phi$ and $\nu$. 
This trend likely arises because increases to the proliferative window width would produce higher rates of overall growth if not compensated for by higher death rates.

Posterior predictions for the medians of the Ki-67 and tumour doubling time observation distributions are presented in Fig. \ref{fig: population_posterior_predictions}. 
These plots confirm that the data medians lie within the medians predicted by the posterior distribution. 
Hence, the posterior distribution is concentrated about a region in parameter space that is consistent with the data. 
We also use the posterior samples to conduct a local sensitivity analysis by calculating the PCC correlation between each inference parameter and the two outputs. 
The results are presented in Fig. \ref{fig: sensitivity}.
The PCC values show that tumour doubling time is most strongly associated with the maximum proliferation rate, $\tilde{\rho}_\text{max}$ (negatively), and the death rate, $\nu$ (positively), as expected.
However, the Ki-67 value is associated more evenly with $\Delta \phi$ (positively), $\tilde{\rho}_\text{max}$ (negatively), $\nu$ (positively) and $t_s$ (negatively), suggesting a highly coupled dependence on the model parameters.

\subsection{Long-term behaviours} \label{sec: population long-term behaviours}

The inference procedure in Sect. \ref{sec: population model parameter calibration} allows us to estimate all model parameters except for the contact inhibition sensitivity, $\kappa$, which is independent of the exponential growth profile $\hat{m}_\text{exp}(\phi)$. 
We numerically explore the impact of $\kappa$ on model behaviour.
In particular, we show that $\kappa$ resolves the three long-term behaviours described in Sect. \ref{sec: structured population dynamics}.

\begin{figure}
    \centering
    \includegraphics[width=0.7\textwidth]{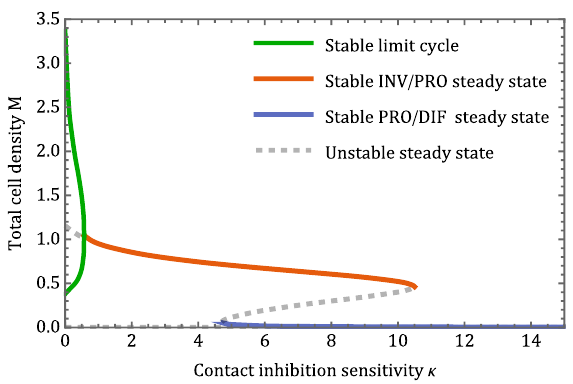}
    \caption{\textbf{Long-term behaviours of the model as a bifurcation diagram with respect to the contact inhibition sensitivity, $\kappa$. } The solid and dashed lines indicate stable and unstable solutions, respectively. A limit cycle emerges from a Hopf bifurcation at $\kappa \approx 0.57$, with upper and lower values of $M(t)$ shown. The INV/PRO steady state exists for $0.57 < \kappa < 10.5$ and is separated from the stable PRO/DIF solution for $\kappa > 4.7$ by a double-fold bifurcation. There are unstable steady states at $M=0$ (corresponding to the exponential growth profile), between the stable steady states and as a continuation of the INV/PRO solution near $\kappa = 0$. Parameter values are set according to the MAP solution \eqref{eqn: population_MAP}. }
    \label{fig: bifurcation}
\end{figure}
Fig. \ref{fig: bifurcation} shows the long-term behaviours of the model as a bifurcation diagram with respect to $\kappa$. 
The plot presents solutions to Eqs. \eqref{eqn: m dynamics}-\eqref{eqn: m nondim} at steady state found using the \textit{FindRoot} routine in Mathematica.
For low values $\kappa < 0.57$, the system exhibits a stable limit cycle.
The green curve represents the upper and lower bounds of the cycle, found numerically by solving the time-dependent equations.
At $\kappa \approx 0.57$, the system undergoes a subcritical Hopf bifurcation that collapses the limit cycle and restores stability to the INV/PRO steady state.
The red INV/PRO steady state exists for an intermediate range $0.57 < \kappa < 10.5$ and is separated from the stable PRO/DIF solution for $\kappa > 4.7$ by a double-fold bifurcation.
Three unstable steady state branches exist;
one at $M = 0$, corresponding to the exponential growth profile, one between the stable steady states and the last is a continuation of the INV/PRO solution near $\kappa = 0$. 

\begin{figure}
    \centering
    \includegraphics[width=0.99\textwidth]{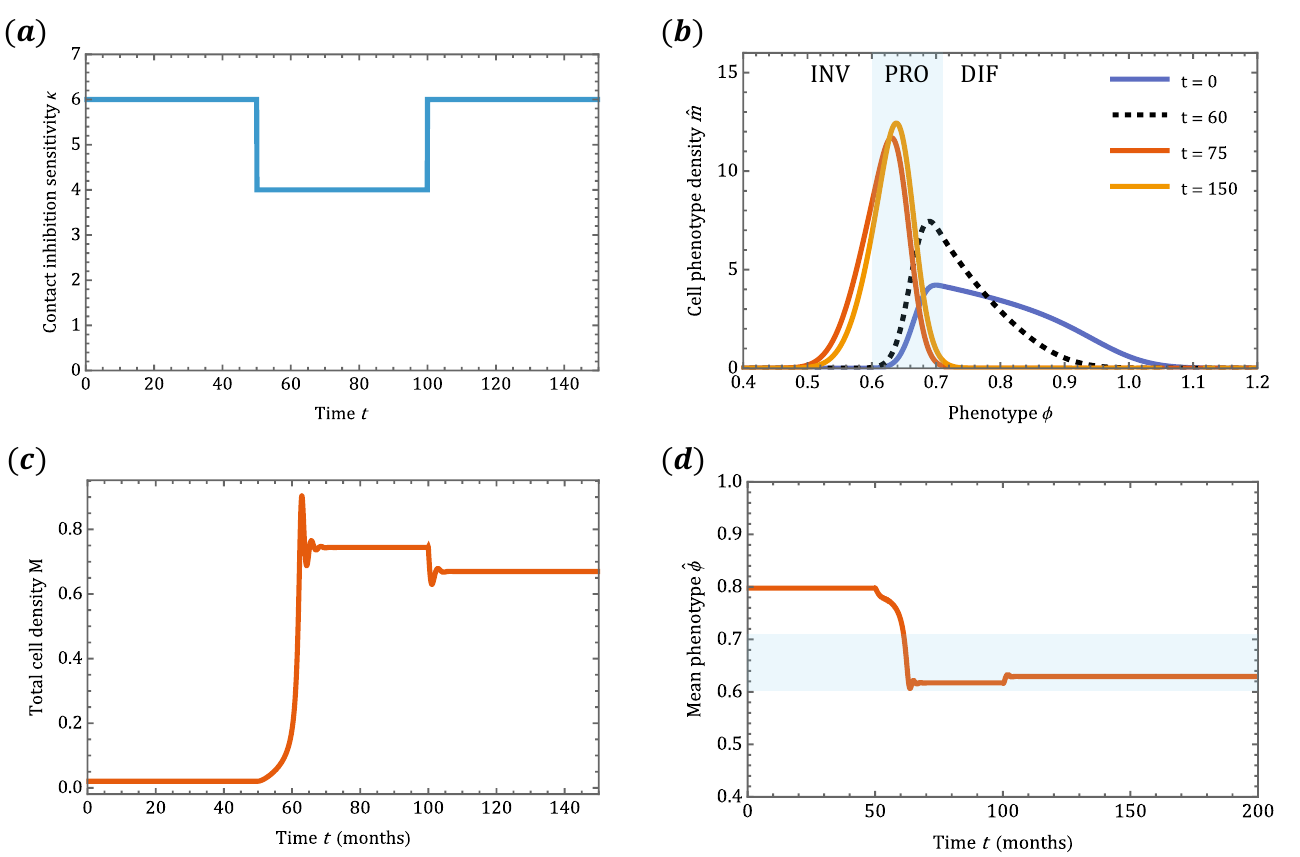}
    \caption{\textbf{Irreversibility of the population phenotype distribution when the contact inhibition sensitivity is decreased and returned to its original value. } The plots show the dynamics of Eqs. \eqref{eqn: m dynamics}-\eqref{eqn: m nondim} when $\kappa$ is decreased from 6 to 4 at $t = 50$ and returned to $6$ at $t = 100$. The initial phenotype distribution is set to the stable PRO/DIF steady state. Plot (a) shows $\kappa$ over time. Plot (b) gives the phenotype distribution at the indicated times. Plots (c) and (d) show the total cell density, $M(t)$, and mean phenotype $\hat{\phi}(t)$. Note that the system remains in the INV/PRO steady state branch even when $\kappa = 6$ is restored. Parameter values are set to the MAP solution \eqref{eqn: population_MAP} with $\lambda_p = 0.35$. }
    \label{fig: irreversibility}
\end{figure}
The existence of the region of bistability in Fig. \ref{fig: bifurcation} indicates that the system should exhibit irreversibility with respect to changes in $\kappa$. 
To test this effect, we conducted a numerical experiment and present results in Fig. \ref{fig: irreversibility}.
The protocol is depicted in Fig. \ref{fig: irreversibility}(a); we lower $\kappa$ from $6$ to $4$ at $t = 50$ and restore its value to $6$ at $t = 100$.
We initialise the system to the stable PRO/DIF steady state at $\kappa = 6$.
Consistent with expectation, the model melanoma does not return to its original PRO/DIF steady state but rather remains on the stable INV/PRO branch. 
Indeed three distinct stable states are evident in the phenotype distribution (Fig. \ref{fig: irreversibility}(b), solid curves), total cell density (Fig. \ref{fig: irreversibility}(c)) and mean phenotype (Fig. \ref{fig: irreversibility}(d)). 
A biological interpretation of the decrease in $\kappa$ at $t = 50$ is the acquisition of mutations that decrease contact inhibition sensitivity (typical in cancer cells \cite{cooper2000development}). 
The results indicate that restoring the degree of contact inhibition to its original value may be insufficient to resolve the phenotype distribution of a tumour.

\section{Spatially resolved population model} \label{sec: spatial}

The structured population model of Sect. \ref{sec: population} neglects spatial effects. 
Tumours exhibit considerable spatial heterogeneity in reality.  
In this section we extend the model to explore the emergence of spatial heterogeneity.
For simplicity, we assume radial symmetry within a two-dimensional geometry to simulate non-metastatic radial growth phase melanoma within the epidermis. 

\subsection{Model statement}

\begin{figure}
    \centering
    \includegraphics[width=0.7\textwidth]{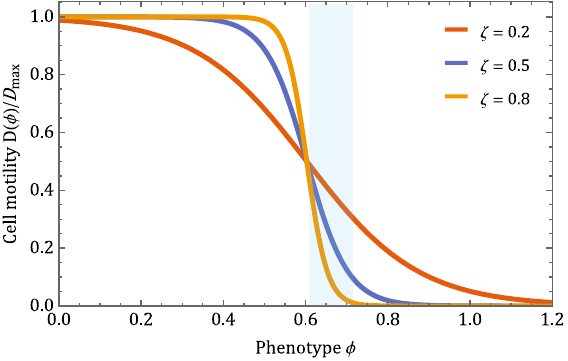}
    \caption{\textbf{Assumed dependence of cell motility on the phenotype variable $\phi$. } The plot illustrates Eq. \eqref{eqn: D(phi)} for $\zeta = 0.2$, $0.5$ and $0.8$. We assume that cell motility decreases monotonically with $\phi$ in a manner centred about $\phi_L$, the invasive end of the proliferative window. The shape parameter $0 < \zeta < 1$ is twice the proportional decrease within $\phi_L < \phi < \frac{\phi_L+\phi_R}{2}$. We set $\phi_L = 0.603$ and $\phi_R = 0.713$, consistent with the MAP solution \eqref{eqn: population_MAP}. }
    \label{fig: spatial_diffusivity} 
\end{figure}

Let $m(\phi, r, t) \geq 0$ be the density of melanoma cells with phenotype $\phi \geq 0$ and distance $r \geq 0$ from the origin at time $t \geq 0$.
We posit that the dynamics are given by
\begin{align}
    \frac{\partial m}{\partial t} + \underbrace{\frac{\partial}{\partial \phi} \bigg[ v m - \frac{\partial}{\partial \phi} (\hat{D} m) \bigg]}_{\text{phenotype modulation}} &= \underbrace{D(\phi) \nabla_r^2 m}_\text{random cell motion} + \big[ \underbrace{\rho(\phi, r,t)}_{\text{proliferation}} - \underbrace{{\color{white}{\rho}}\nu {\color{white}{\rho}}}_{\text{death}} \big] m, \label{eqn: m spatial}
\end{align}
where $\nabla_r^2 = \frac{\partial^2}{\partial r^2} + \frac{1}{r} \frac{\partial}{\partial r}$ is the two-dimensional Laplacian operator under radial symmetry.
We assume the cell motility coefficient $D(\phi)$ depends on phenotype according to
\begin{align}
    &D(\phi) =  \frac{D_\text{max}}{1 + e^{Q(\phi - \phi_L)}}, & &Q = \frac{4 \, \text{arctanh}(\zeta)}{\phi_R - \phi_L}. \label{eqn: D(phi)}
\end{align}
The functional form \eqref{eqn: D(phi)} is monotone decreasing and centred about $\phi_L$ such that $D(\phi_L) = D_\text{max}/2$.
This shape ensures that cells experience their greatest increase in motility as they exit the invasive end of the proliferative window.
The shape parameter $0 < \zeta < 1$ is twice the proportional decrease of $D(\phi)$ within $\phi_L < \phi < \frac{\phi_L+\phi_R}{2}$.
That $D(\phi)$ is decreasing reflects the MITF rheostat; MITF-low invasive cells are highly motile relative to differentiated MITF-high cells \cite{goding2011picture, hsiao2014roles}.
A plot of Eq. \eqref{eqn: D(phi)} is given in Fig. \ref{fig: spatial_diffusivity}. 
The remaining terms are spatially resolved counterparts to their definitions in Sect. \ref{sec: population}. 
We define them explicitly and non-dimensionalise the model in Appendix \ref{Appendix: spatial}. \\

\noindent \textbf{Boundary conditions. } We impose no-flux boundary conditions in phenotype and space
\begin{align}
    &\bigg[ v m - \frac{\partial}{\partial \phi} (\hat{D}m) \bigg]_{\phi \rightarrow 0, \infty} = 0, & &\frac{\partial m}{\partial r}\bigg\vert_{r \rightarrow 0, \infty} = 0 \label{eqn: m spatial noflux},
\end{align}
so that cells may not leave the domain via unrealistic fluxes.  \\

\noindent \textbf{Initial conditions. } We close the dimensionless model with initial conditions
\begin{align}
    &m(\phi, r, 0) = M_0 \cdot \hat{m}_0 (\phi) \cdot H(r - r_0), & &M_0 \ll 1,& &\int_0^\infty \hat{m}_0(\phi) d\phi = 1, \label{eqn: m spatial init}
\end{align}
where $H(\cdot)$ is the Heaviside step function.
For numerical solutions we take $r_0 = 0.1$ (mm) and $M_0 = 10^{-3}$ so that the tumour begins with a small number of cells in a neighbourhood of the origin.
To avoid early transients we also assume $\hat{m}_0(\phi)$ satisfies the balanced exponential growth Eqs. \eqref{eqn: exp growth pde}-\eqref{eqn: v,D,rho tildes}. \\

\noindent \textbf{Parameter values. } We do not pursue a formal parameter inference for the spatial model due to the scarcity of spatio-temporal data for the horizontal growth phase. 
We note that the motility coefficient $D_\text{max}$ may be scaled out by using a length scale proportional to $\sqrt{D_\text{max} \cdot [1 \, \text{month}]}$. 
In Sect. \ref{sec: spatial numerics}, we fix $D_\text{max} = 0.003 \, \text{mm}^2 /\text{month}$ for illustration as this yields growth speeds consistent with reported values of the order $0.1 \,\text{mm}/\text{month}$  \cite{liu2006rate, beer2011growth}. 
The remaining parameters are fixed to their MAP estimates \eqref{eqn: population_MAP}, except $\kappa$ and $\zeta$, which we explore across a range of values.

\subsection{Numerical spatio-temporal solutions} \label{sec: spatial numerics}

We generate numerical solutions to the spatially resolved model \eqref{eqn: m spatial}-\eqref{eqn: m spatial init} using the method of lines. 
Specifically, we use a uniform second-order central differencing in $\phi$ and $r$ across a computational domain $0.2 < \phi < 1.4$  and $0 < r < 3 \,$ mm with $150$ phenotype nodes and $120$ spatial nodes. 
Integrals are approximated with the trapezoidal rule and the resulting system of ODEs solved via the \textit{NDSolve} routine in Wolfram Mathematica.

\begin{figure}
    \centering
    \includegraphics[width=0.99\textwidth]{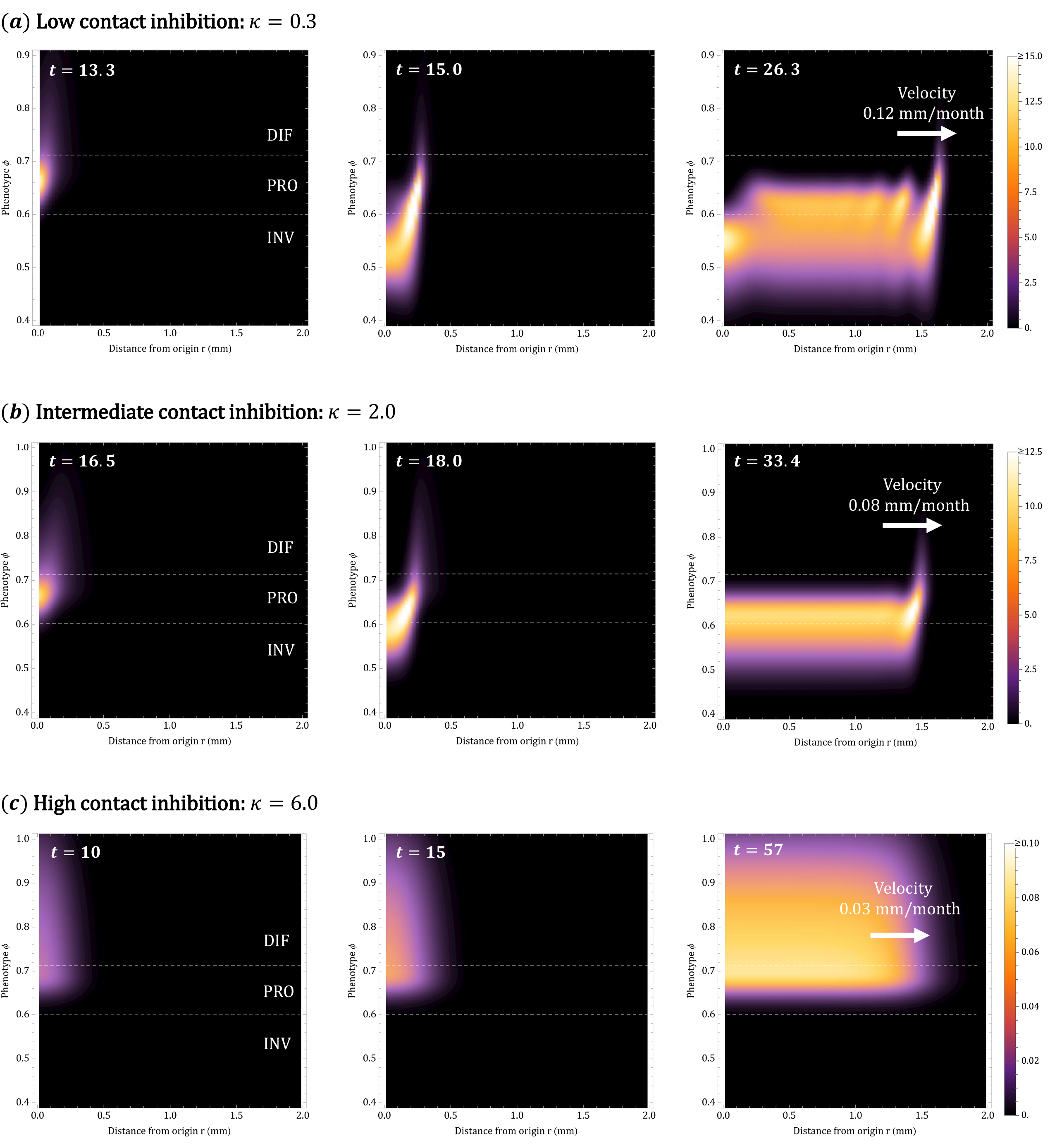}
    \caption{\textbf{The model melanoma propagates as a wave with core behaviour that depends on the contact inhibition sensitivity $\kappa$. } The plots above show the density $m(\phi, r, t)$ at the indicated times for $\kappa = 0.3$, $2.0$ and $6.0$. (a) When contact inhibition sensitivity is low the core $r = 0$ oscillates and generates a series of density peaks about the wavefront. (b) Intermediate sensitivity to contact inhibition produce model melanomas in which the core settles to a spatially homogeneous INV/PRO steady state. (c) High sensitivity to contact inhibition yield a slow wave with PRO/DIF core and wavefront.
    The proliferative window is indicated by the dashed lines. We set $\zeta = 0.5$.}
    \label{fig: m(phi,r,t)}
\end{figure}
Solutions for the density $m(\phi, r, t)$ are presented in Fig. \ref{fig: m(phi,r,t)}. 
We find the model melanoma exits the exponential growth phase in a neighbourhood of $r = 0$ and grows into a wave-like solution that propagates at constant speed as $t \rightarrow \infty$. 
The behaviour of the core ($r = 0$) depends on the value of $\kappa$ in a manner consistent with the spatially homogeneous bifurcation diagram, Fig. \ref{fig: bifurcation}. 
For low levels of contact inhibition sensitivity (e.g., $\kappa = 0.3$, see Fig. \ref{fig: m(phi,r,t)}(a)) the solution propagates rapidly with multiple density peaks about the wavefront due to oscillations at $r = 0$.
Intermediate sensitivity to contact inhibition (e.g., $\kappa = 2.0$, see Fig. \ref{fig: m(phi,r,t)}(b)) produces solutions with a single density peak near the wavefront and a spatially homogeneous INV/PRO core.
Finally, high sensitivity to contact inhibition (e.g., $\kappa = 6.0$, see Fig. \ref{fig: m(phi,r,t)}(c)) yields a homogeneous PRO/DIF core and slow wavefront. 
The wavefront always consists of a PRO/DIF phenotype distribution, regardless of the value of $\kappa$, due to its vicinity to low cell densities.
Invasive phenotypes $\phi < \phi_L$ are found closely behind the wavefront and in the core for $\kappa = 0.3$ and $2.0$.

\begin{figure}
    \centering
    \includegraphics[width=0.8\linewidth]{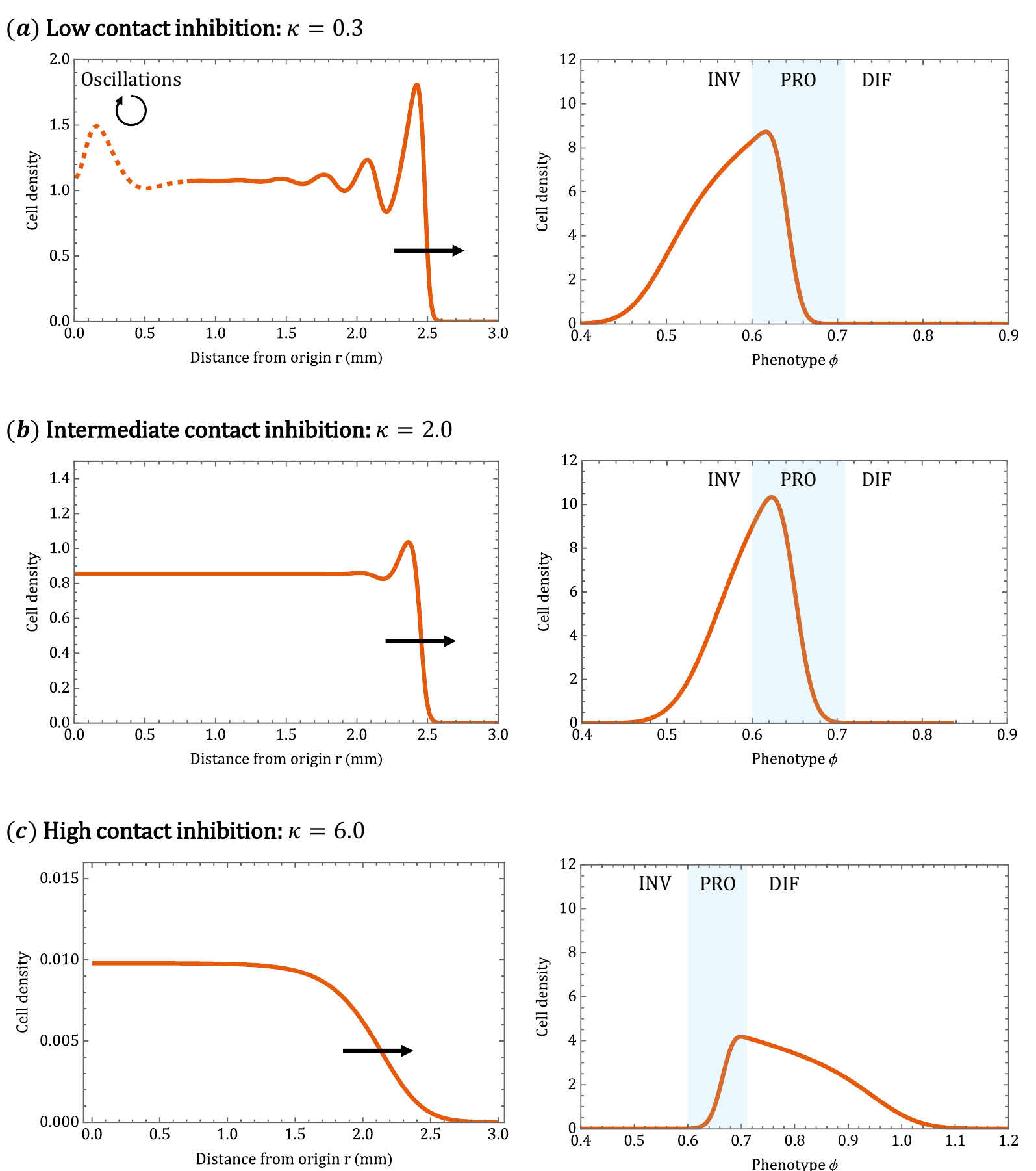}
    \caption{\textbf{Model solutions exhibit various spatial and phenotype profiles during late-stage wave propagation. } 
    The marginal spatial and phenotype distributions are shown on the left and right, respectively, for (a) $\kappa = 0.3$, (b) $\kappa = 2.0$ and (c) $\kappa = 6.0$.  
    By contrast to the spatially homogeneous model, all solutions yield a steady state phenotype distribution as $t \rightarrow \infty$. We set $\zeta = 0.5$. }
    \label{fig: spatial_latestage}
\end{figure}

The marginal spatial and phenotype distributions during late-stage wave propagation are shown in Fig. \ref{fig: spatial_latestage}. 
The spatial profile in Fig. \ref{fig: spatial_latestage}(a) shows that the oscillatory core for low $\kappa$ neighbours a region of spatial homogeneity. 
As $r$ increases, this region exhibits a progression of peaks and troughs of increasing amplitude until the largest peak at the wavefront.
As $t$ increases, the oscillatory region remains fixed in size while the spatially homogeneous region grows with the addition of smaller and smaller peaks.
Consequently, and in contrast to the spatially homogeneous model, the phenotype distribution settles to a steady state as $t \rightarrow \infty$.
Interestingly, the phenotype distribution matches the (unstable) steady state distribution when $\kappa = 0.3$ in the model of Sect. \ref{sec: population}; cell motion renders the distribution stable. 
Late-stage solutions for $\kappa = 2.0$ and $\kappa = 6.0$, shown in Fig. \ref{fig: spatial_latestage}(b) and Fig. \ref{fig: spatial_latestage}(c) respectively, exhibit spatially homogeneous cores with phenotype distributions consistent with their corresponding profiles in Fig. \ref{fig: 3longterm_behaviours}. 
The wavefront for high contact inhibition lacks a density peak and has a more gradual decrease to zero as $r \rightarrow \infty$. 

\begin{figure}
    \centering
    \includegraphics[width=0.99\textwidth]{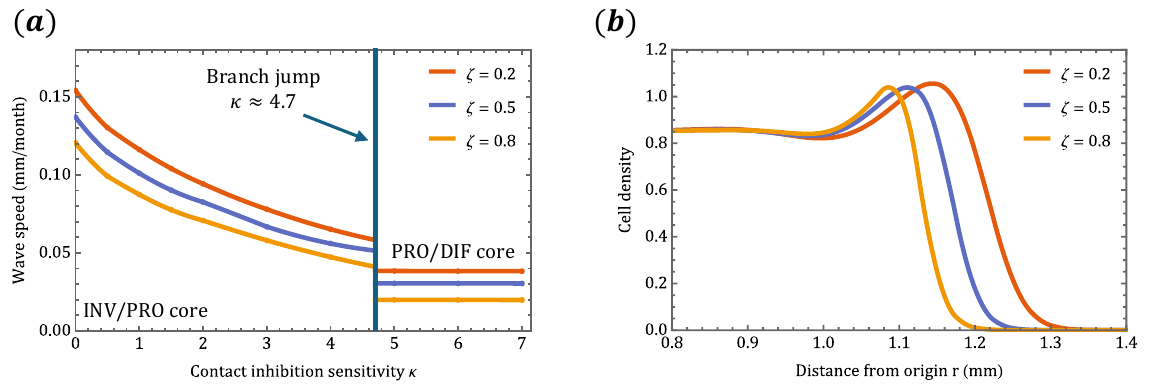}
        \caption{\textbf{The free parameters $\kappa$ and $\zeta$ influence late-stage wave propagation. } (a) Wave speed decreases with both $\kappa$ and $\zeta$. Solutions to Eqs. \eqref{eqn: m spatial}-\eqref{eqn: m spatial init} are attracted to different branches depending on $\kappa$ in a manner consistent with Fig. \ref{fig: bifurcation}. (b) Increases to the motility shape parameter $\zeta$ sharpen the wavefront, here evaluated for $\kappa = 2$ at $t = 30$. }
    \label{fig: sweep}
\end{figure}
We explore the impact of the free parameters $\kappa$ and $\zeta$ on late-stage wave propagation using a parameter sweep. 
Key results are shown in Fig. \ref{fig: sweep}. 
We compute the wave speed by numerically differentiating the location of the front, defined here as the largest value of $r$ for which $M(r,t) = \sup_{r \geq 0} M(r,t)/2$.    
The time derivative is taken once $t$ is sufficiently large that the wave speed is approximately constant (typically $t = 30$).
The results, given in Fig. \ref{fig: sweep}(a), indicate that wave speed decreases monotonically with both $\kappa$ and $\zeta$. 
These trends are consistent with a wave that is pulled by proliferation near the front rather than pushed by nonlinear dynamics in the core.
Indeed, higher values of $\kappa$ yield lower rates of proliferation and higher values of $\zeta$ reduce the motility of proliferative cells with $\phi_L < \phi < \phi_R$ (recall Fig. \ref{fig: spatial_diffusivity}). 
The results also confirm that solutions to Eqs. \eqref{eqn: m spatial}-\eqref{eqn: m spatial init} are attracted to different branches depending on $\kappa$ in a manner consistent with Fig. \ref{fig: bifurcation}.
In particular, values $\kappa > 4.7$ yield slower waves of the type shown in Fig. \ref{fig: spatial_latestage}(c). 
Lastly, we find that increases to the shape parameter $\zeta$ sharpen the wavefront (see Fig. \ref{fig: sweep}(b)).
This trend reflects the motility reduction as $\zeta$ is increased for cells with higher phenotype values $\phi > \phi_L$.  


\section{Discussion} \label{sec: discussion}

In this paper we have developed a new mathematical model to investigate the emergent cell population dynamics of the melanoma MITF rheostat. 
The model is formulated as a multiscale and phenotype-structured partial differential equation for melanoma cell density in the epidermis.
We analysed the model in three stages, first considering subcellular MITF dynamics 
then well-mixed population dynamics representative of the tumour core 
and, finally, spatially resolved, radially-symmetric population dynamics. 
The model was kept 
deliberately simple in order to calibrate parameter values to published data using Bayesian inference. 
Below we summarise the key findings and discuss their biological and mathematical implications. \\

\noindent \textbf{Subcellular model. }\\
We modelled the subcellular MITF dynamics as a stochastic system of differential equations for RNA concentration, protein concentration and phenotype. 
The results highlight the heteroscedasticity evident in single-cell sequencing data, where the RNA variance increases with its mean. 
We quantified this effect using a power-law stochastic term in the RNA dynamics \eqref{eqn: r}.
Some of this variance may arise from population selection effects or spatial heterogeneity, rather than subcellular noise.
However, subcellular RNA dynamics occur on a shorter timescale than population selection (MITF RNA has a half-life of approximately 2.5 hours, whereas cell proliferation and turnover occur over days to months), suggesting 
that subcellular noise is the dominant contribution. 
Moreover, similar heteroscedasticity has been observed in melanoma cell lines for which spatial heterogeneity is minimal \cite{wouters2020robust}. 

From a biological perspective, the results of Sect. \ref{sec: subcellular} highlight an important nuance in the context of the MITF rheostat. 
{\color{black}Since MITF expression is suppressed by the integrated stress response, a more stressful microenvironment produces cells with lower RNA heterogeneity and, consequently, less variance in a phenotype variable 
downstream of RNA. 
Indeed, lower transcription rates select for narrower subcellular phenotype distributions within individual cells in our model (recall Fig. \ref{fig: phenotype_gaussians}). 
This result has two possible interpretations. 
MITF-low cells may be more functionally similar to each other than MITF-high cells.
Alternatively, features relevant to an MITF-low state, such as cell motility, may depend more sensitively on MITF than features of an MITF-high state, such as pigmentation.}

From a mathematical perspective, the derivation of the phenotype flux, detailed in Appendix \ref{Appendix: phenotype flux}, may serve as a template for incorporating single-cell RNA sequencing data into the phenotype flux of future structured population models.
The derivation follows standard multiscale homogenisation arguments for fast–slow stochastic systems \cite{PavliotisStuart2008} and is based on a separation of timescales between RNA and protein dynamics and a slower phenotype variable. 
Although reduction to a single variable is not strictly necessary when considering the dynamics of individual cells, the procedure considerably reduces the computational burden of simulating multiple structure variables at the population scale. 
A useful outcome of homogenisation is that phenotypic drift and diffusion are expressed in terms of measurable subcellular parameters.
By contrast, previous structured population models have introduced diffusion either as an independent parameter or via the upscaling of a discretised structure variable \cite{lorenzi2022trade, chambers2023new, agostinelli2026multiscale}.
\\

\noindent \textbf{Population model. }\\
In Sect. \ref{sec: population}, we studied spatially homogeneous, phenotype-resolved melanoma cell population dynamics, in which the average MITF transcription rate is a decreasing function of the total cell density.
Here, cell density serves as a proxy for cell stress, reflecting higher nutrient competition, reduced oxygen availability and/or mechanical confinement, 
the latter of which has recently been shown to drive invasive gene programs \cite{hunter2025mechanical}. 
After an initial transient, the model predicts a prolonged period of balanced exponential growth in a population dominated by proliferative cells and differentiated non-cycling cells.
Once the cell density is high enough to affect MITF expression, the system evolves to one of three stable behaviours depending on cell sensitivity to contact inhibition, $\kappa$.
For high $\kappa$, there is a low-density ``PRO/DIF" steady state balance between proliferative (PRO) cells and non-cycling differentiated (DIF) cells.
Intermediate values of $\kappa$ yield a high density ``INV/PRO" steady state balance between non-cycling invasive cells (INV) and proliferative cells.
Finally, low values of $\kappa$ produce a cyclic solution that is periodically dominated by either INV or PRO cells. 

It is tempting to identify the PRO/DIF state with a healthy melanocyte population.
However, melanocytes are typically longer-lived (e.g., years  rather than months \cite{cichorek2013skin}) than suggested by the melanoma TUNEL-derived timescale (Appendix \ref{Appendix: TUNEL}). 
We therefore interpret the PRO/DIF state either as a non-cancerous but atypical melanocyte population with increased turnover, or as a superficial spreading melanoma that would require further genetic mutations to become metastatic. 
By contrast, the INV/PRO state is sustained by stress-induced switching to invasive phenotypes at high cell densities, consistent with nodular melanoma and superficial spreading melanoma that has transitioned to vertical growth \cite{greenwald2012superficial}. 
The cyclic state predicted for very low contact inhibition sensitivity is harder to reconcile with biological observations. 
The density oscillations differ from previously reported cell cycle synchrony in melanoma cell lines \cite{gavagnin2021synchronized}.
In our model, cell density can vary five-fold over approximately two months when $\kappa = 0$.
Although the spatial model suggests oscillations may be confined to the tumour core, changes of this magnitude would likely have been reported clinically.
We therefore regard the cyclic solutions as biologically unrealistic and infer a lower bound of $\kappa>0.57$ for the model parameters.

The population model analysis revealed an important nuance concerning phenotype reversibility.
While the MITF rheostat permits reversible phenotype switching at the single-cell level, the bistability between the PRO/DIF and INV/PRO states gives rise to hysteresis at the population level. As shown in Fig. \ref{fig: irreversibility}, a population initially in the PRO/DIF state can switch to the INV/PRO state when contact inhibition sensitivity is sufficiently reduced, for example by mutation. However, restoring sensitivity to its previous level does not return the population to the PRO/DIF state. Thus, reversibility at the cellular level does not imply reversibility at the population scale.
Because the behaviour of individual cells is coupled via the cell density, 
population level behaviour cannot, in general, be inferred by treating the cells as independent and aggregating their dynamics. \\

\noindent \textbf{Spatially resolved population model. } \\
Finally, in Sect. \ref{sec: spatial}, we extended the population model to a spatially-resolved, radially symmetric setting. Cell movement was assumed to be random, with phenotype-dependent motility so that MITF-high differentiated cells are less motile than MITF-low invasive cells. Simulations generated melanomas that expand as travelling waves, ultimately propagating at a constant radial speed. The behaviour of the tumour core is consistent with the predictions of the spatially homogeneous model and, in particular, depends on sensitivity to contact inhibition. High sensitivity yields a slowly propagating PRO/DIF state, while intermediate levels generate a faster wave with a PRO/DIF wavefront and INV/PRO core. For low sensitivity, the wave speed increases further and the tumour core exhibits sustained oscillations. Analysis of the wave speed indicated that expansion is driven primarily by proliferation at the leading edge, consistent with a pulled front rather than one pushed by nonlinear dynamics within the core.


The presence of a PRO/DIF wavefront in all solutions is a direct consequence of using cell density as a proxy for stress. Cells at the tumour edge experience the lowest densities and, therefore, the highest MITF transcription rates. While cell density is a plausible inverse measure of stress within the epidermis, it is unlikely to provide a complete description of the microenvironment. In particular, vertical-growth-phase melanomas typically exhibit an invasive front in the dermis \cite{maiques2025matrix}, suggesting that additional environmental factors would be required to reproduce this behaviour.


The oscillating core solution may be of further mathematical interest. 
The oscillating core borders a spatially homogeneous region whose phenotype distribution matches the unstable INV/PRO branch observed at low $\kappa$. This observation suggests that cell movement suppresses oscillations and stabilises the INV/PRO state throughout much of the tumour bulk, while oscillatory behaviour persists only near the core.
{\color{black}We note that similar patterns have been reported in the travelling waves of cyclic predator-prey systems (e.g., Fig. 6 in \cite{sherratt2008periodic}).}
\\


\noindent \textbf{Future work. } \\
The model framework is readily extended to study the vertical growth phase of melanoma. 
The main challenges are to identify an appropriate measure of cell stress in the dermis and couple it to MITF expression.
To generate an invasive front, it may be necessary to explicitly account for extracellular matrix.
Moreover, we are interested in the impact of various cells in the melanoma microenvironment.
While early non-invasive melanomas are primarily surrounded by keratinocytes, as the tumour penetrates into deeper subcutaneous tissue, melanoma cells become surrounded by lipid-laden adipocytes.
Adipocytes participate in fatty acid signalling with melanoma cells that may stabilise the invasive phenotype \cite{zhang2018adipocyte, chocarro2025fatty}. 
Resolving these contributions by extending the current model is a promising direction of ongoing work. 

The subcellular dynamics 
may also be adjusted to account for possible subcellular bistability.
Several positive-feedback loops may generate bistability between MITF-low invasive and MITF-high differentiated states.
Examples include feedback involving MITF and BRN2 \cite{goodall2008brn}, fatty acid saturation \cite{vivas2020lineage} and components of the gene regulatory network in \cite{Subhadarshini2023PDL1Heterogeneity}.
Understanding the population-level impact of subcellular phenotypic bistability is the subject of ongoing work. \\

\noindent \textbf{Conclusions. } \\
The mathematical model developed in this paper shows that the MITF rheostat can generate three stable long-term population behaviours in the epidermis. 
In order of decreasing sensitivity to contact inhibition, these are a slow-growing melanoma of proliferative cells and non-cycling differentiated cells without invasive potential, a faster-propagating melanoma with a steady state invasive core, and a rapidly growing melanoma with oscillatory core dynamics. 
More broadly, the analysis highlights that phenotype reversibility at the level of individual cells does not imply reversibility of phenotype distributions at the population scale.
Single-cell properties, such as the reversibility of invasive capacity, must therefore be extrapolated with caution to populations with coupled cell dynamics.
These findings further the understanding of melanoma cell populations and cell plasticity more generally.

\backmatter

\subsection*{Acknowledgements}
All authors were supported by the Ludwig Institute for Cancer Research (Oxford branch). KLC thanks Jeremy H. Raymond for many insightful conversations on melanoma and Seemadri Subhadarshini for helpful comments on the manuscript.


\subsection*{CRediT authorship contribution statement}
\textbf{KLC:} Writing– original draft, Formal analysis, Software, Methodology, Conceptualisation, Investigation, Data curation, Visualisation. 
\textbf{RMW:} Writing– review and editing, Conceptualisation, Supervision. 
\textbf{CRG:} Writing– review and editing, Conceptualisation, Supervision. \textbf{HMB:} Writing– review and editing, Conceptualisation, Methodology, Validation, Supervision.

\subsection*{Competing interests}
The authors declare no competing interests.

\subsection*{Data availability}
No new biological or clinical data were generated in this study.
The single-cell RNA-sequencing dataset analysed in this study is publicly available from the Gene Expression Omnibus under accession code GSE72056.
The remaining experimental measurements used for model calibration were digitised from the published studies cited in the manuscript.
The digitised values and data underlying the figures are available in the GitHub repository \url{https://github.com/KL-Chambers/MITF_rheostat_population_dynamics}.

\subsection*{Code availability}

All analyses in this study used the software Mathematica (version 14.2). The code is also available publicly at \url{https://github.com/KL-Chambers/MITF_rheostat_population_dynamics}.

\bibliography{sn-bibliography}


\newpage
\counterwithin{figure}{section}
\counterwithin{table}{section}
\appendix

\section{MCMC plots: subcellular inference problem} \label{Appendix: subcellular_MCMC}

\begin{figure}[h]
    \centering
    \includegraphics[width=0.99\textwidth]{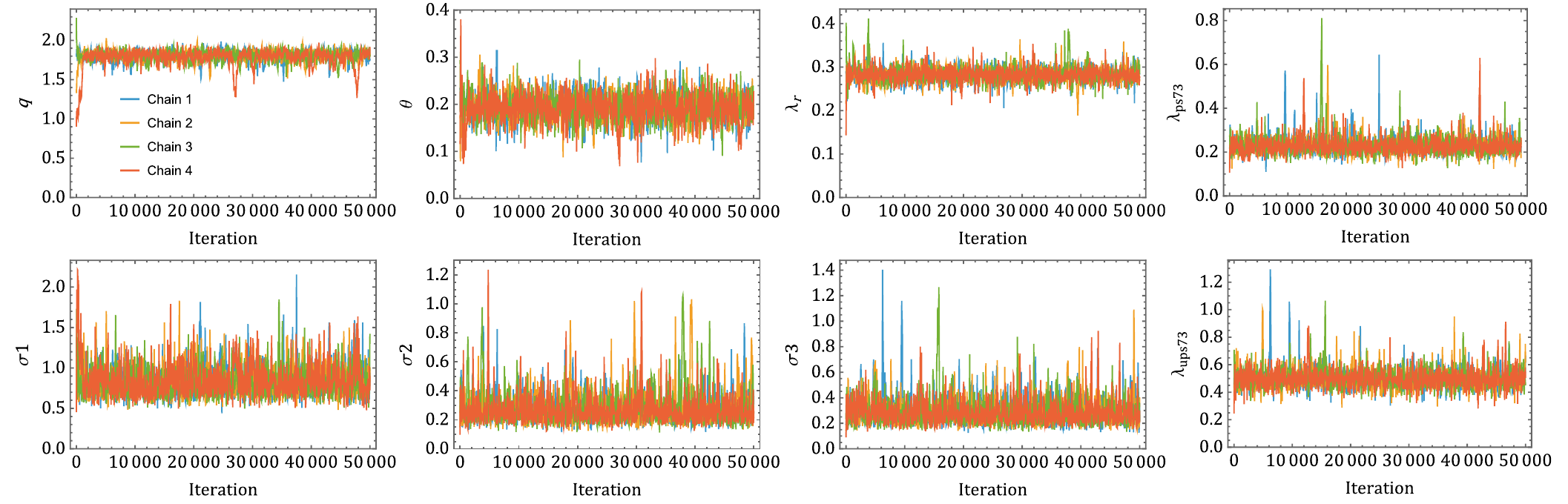}
    \caption{\textbf{Trace plots of the four independent chains.} The chains mix well and remain stationary. The Gelman-Rubin statistic is less than $1.01$ for each parameter when the first $10000$ iterations are discarded as a burn-in period.}
    \label{fig: subcellular_MCMC_chains}
\end{figure}

\begin{figure}[h]
    \centering
    \includegraphics[width=0.99\textwidth]{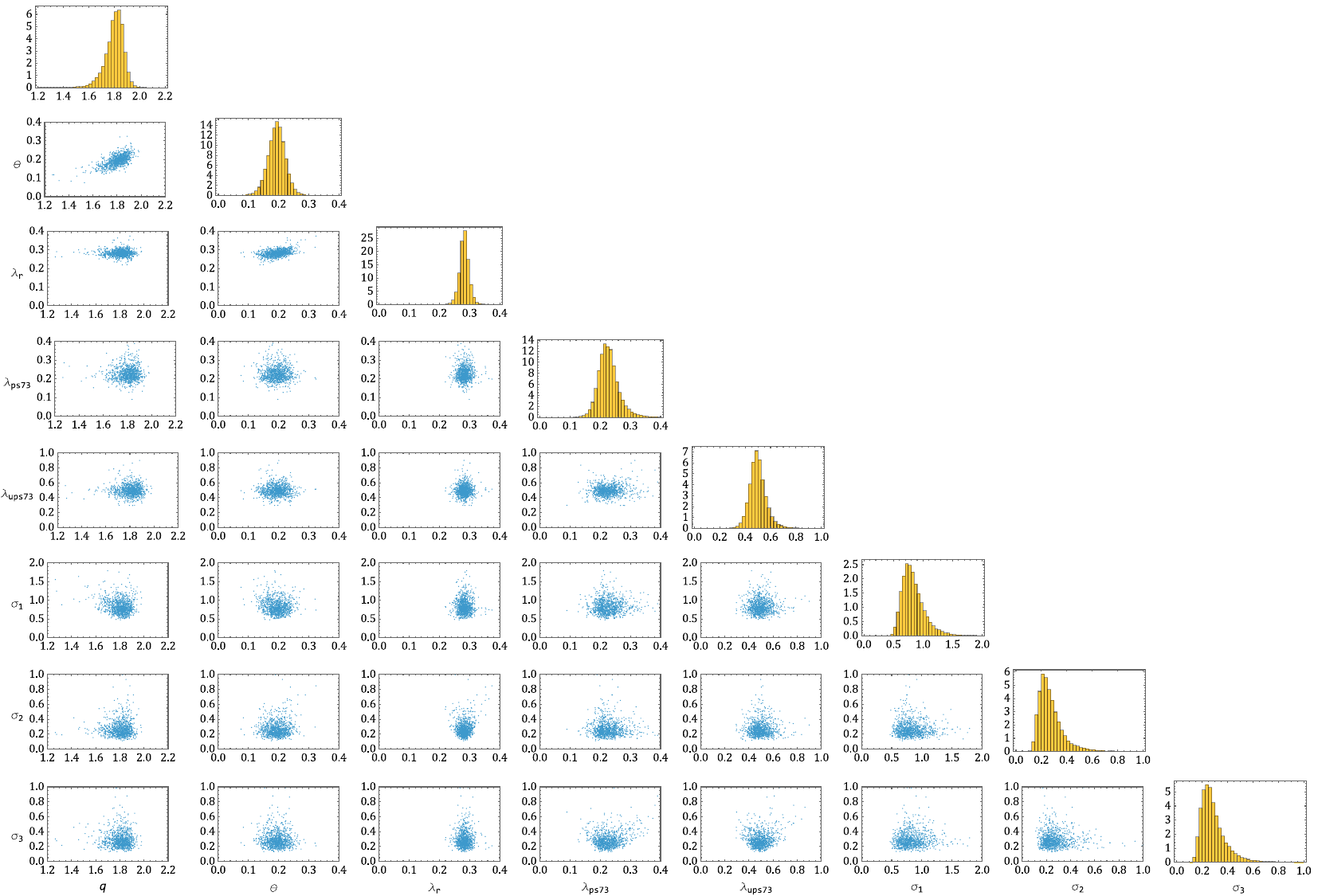}
    \caption{\textbf{Posterior samples for all parameters in the subcellular inference problem. } We note that the observation noise parameters $\sigma_i$, $i= 1$, 2, 3 are also practically identifiable and uncorrelated with the other parameters of the model. }
    \label{fig: subcellular_MCMC_full}
\end{figure}

\begin{figure}[h]
    \centering
    \includegraphics[width=0.99\textwidth]{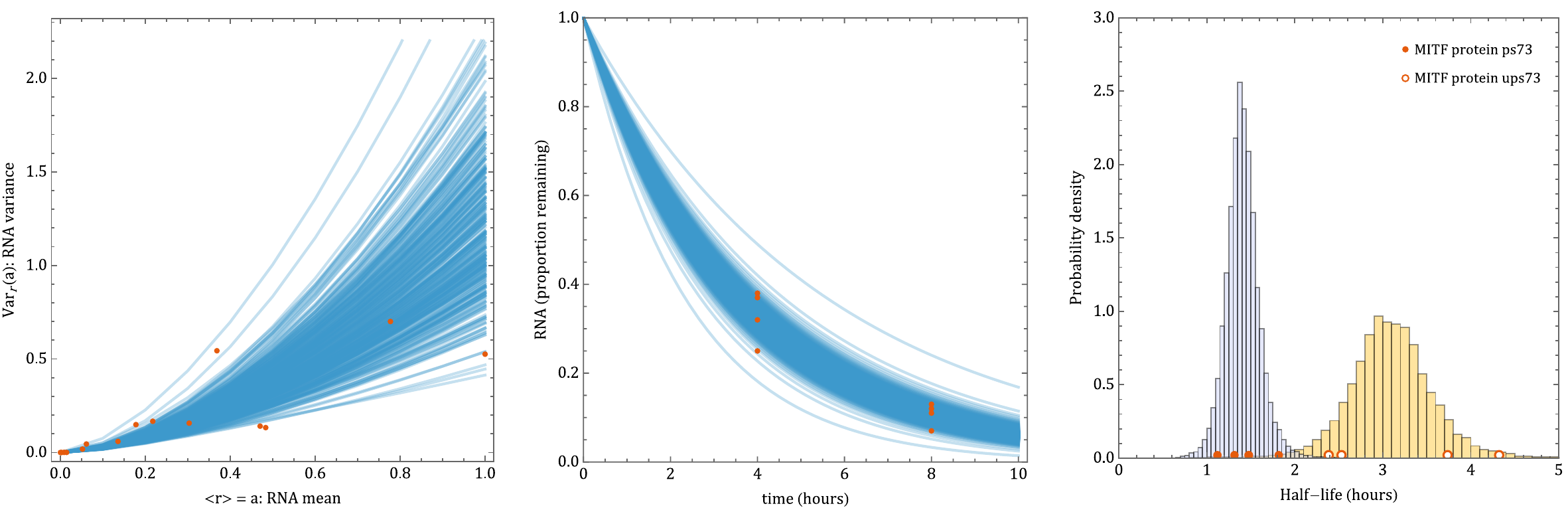}
    \caption{\textbf{Posterior predictions of the true (observation noise-free) outputs. }The predictions for the RNA mean-variance relationship, RNA exponential decay and protein half-lives cover the data well. }
    \label{fig: subcellular_MCMC_PPC}
\end{figure}

\clearpage
\section{Derivation of the phenotype flux} \label{Appendix: phenotype flux}

We expand the probability density $\psi(r,p,\phi,t)$ and the phenotype marginal distribution $\Psi (\phi, t)$ as an asymptotic series of the form
\begin{align}
    &\psi = \psi_0 + \epsilon \psi_1 + \dots, & &\Psi = \Psi_0 + \epsilon \Psi_1 + \dots . \label{eqn: psi series}
\end{align}
From the normalisation condition $\int_0^\infty \Psi d\phi = 1$, the functions $\Psi_i$ satisfy
\begin{align}
    &\int_0^\infty \Psi_0 d\phi = 1, & &\int_0^\infty \Psi_i d\phi = 0 \text{ for } i=1,2, \dots .
\end{align}

\noindent \textbf{Balance at $\mathcal{O}(1)$. } Substituting Eq. \eqref{eqn: psi series} into Eq. \eqref{eqn: fokker-Planck psi} and equating the $\mathcal{O}(1)$ terms yields
\begin{align}
    \mathcal{L}_0 \psi_0 = 0.
\end{align}
Using that $\mathcal{L}_0$ is independent of time derivatives from Eq. \eqref{eqn: L0}, we deduce that 
\begin{align}
    &\psi_0(r,p,\phi,t) = \pi(r,p, t) \Psi_0(\phi, t), \label{eqn: psi0}
\end{align}
for a function $\pi$ satisfying $\mathcal{L}_0 \pi = 0$ and $\int_0^\infty \int_0^\infty \pi drdp = 1$. The notation $\pi$ is chosen deliberately since $\mathcal{L}_0 \pi = 0$ is identical to the earlier equilibrium condition \eqref{eqn: pi}. Both objects represent the quasi-steady probability distribution of $r$ and $p$ on the slow timescale and so we retain the equilibrium analysis of $\pi$ in Sect. \ref{sec: rp equilibrium}. \\

\noindent \textbf{Balance at $\mathcal{O}(\epsilon)$. } Substituting Eq. \eqref{eqn: psi series} into Eq. \eqref{eqn: fokker-Planck psi} and equating the $\mathcal{O}(\epsilon)$ terms gives
\begin{align}
    \partial_t \psi_0 + \mathcal{L}_0 \psi_1 + \mathcal{L}_1\psi_0 = 0. \label{eqn: O(e) balance}
\end{align}
Hence, using that $\psi_0 = \pi (r,p,t) \Psi_0(\phi,t)$, we find
\begin{align}
    \partial_t (\pi \Psi_0) + \mathcal{L}_0 \psi_1 + \gamma \partial_\phi \big[ (p - \phi) \pi \Psi_0 \big] = 0. \label{eqn: O(e) explicit}
\end{align}
Integrating Eq. \eqref{eqn: O(e) explicit} with respect to $r$ and $p$ yields
\begin{align}
    \partial_t \Psi_0 + \int_0^\infty \int_0^\infty \mathcal{L}_0 \psi_1 drdp + \gamma \partial_\phi \langle (p - \phi) \Psi_0 \rangle_\pi = 0.
\end{align}
The integral term vanishes due to the $\mathcal{O}(\epsilon)$ balance of the no-flux boundary conditions \eqref{eqn: psi bconds} and the third term simplifies using the result $\langle p \rangle_\pi = a$ from Eq. \eqref{eqn: pi moments}. Hence, we obtain
\begin{align}
    \partial_t \Psi_0 + \gamma \partial_\phi [ (a - \phi) \Psi_0] = 0. \label{eqn: f0 result}
\end{align}
That is, the $\phi$-marginal distribution satisfies an advection equation at leading order with a drift velocity $\gamma(a-\phi)$ that drives $\phi$ towards the common mean value, $a(t)$, of the quasi-steady RNA and protein concentrations. \\

\noindent \textbf{Balance at $\mathcal{O}(\epsilon^2)$. } Equating the $\mathcal{O}(\epsilon^2)$ terms in Eq. \eqref{eqn: fokker-Planck psi} gives
\begin{align}
    \partial_t \psi_1 + \mathcal{L}_0 \psi_2 + \mathcal{L}_1 \psi_1 = 0, \label{eqn: O(e^2) balance}
\end{align}
which, upon integrating with respect to $r$ and $p$, becomes
\begin{align}
    \partial_t \Psi_1 + \int_0^\infty \int_0^\infty \mathcal{L}_0 \psi_2 drdp + \int_0^\infty \int_0^\infty \mathcal{L}_1 \psi_1 drdp = 0. \label{eqn: f1 pde}
\end{align}
The first integral vanishes due to the $\mathcal{O}(\epsilon^2)$ balance of the no-flux boundary conditions \eqref{eqn: psi bconds}. To evaluate the second integral, we write
\begin{align}
    \int_0^\infty \int_0^\infty \mathcal{L}_1 \psi_1 drdp  &= \gamma \partial_\phi \int_0^\infty \int_0^\infty (p  -  \phi) \psi_1 drdp, \nonumber \\
    &= \gamma \partial_\phi \int_0^\infty \int_0^\infty \big[ (p - a) + (a - \phi) \big] \psi_1 drdp \nonumber \\
    &= \gamma \partial_\phi \int_0^\infty \int_0^\infty (p-a) \psi_1 drdp + \gamma \partial_\phi \big[ (a - \phi) \Psi_1 \big].
\end{align}
To evaluate the remaining integral, it is convenient to introduce the auxiliary function
\begin{align}
    \chi (r,p) = \frac{r}{\lambda_r} + \frac{p}{\lambda_p}, \label{eqn: chi def}
\end{align}
so that we may write
\begin{align}
   \int_0^\infty \int_0^\infty (p-a) \psi_1 drdp &=  \int_0^\infty \int_0^\infty \chi \mathcal{L}_0 \psi_1 drdp \quad \text{(integration by parts)} \nonumber \\
    &= -\int_0^\infty \int_0^\infty \chi (\partial_t + \mathcal{L}_1 )\psi_0 drdp \quad \text{(Eq. \eqref{eqn: O(e) balance})} \nonumber \\
    &= - \int_0^\infty \int_0^\infty \chi (\partial_t + \mathcal{L}_1) (\pi \Psi_0) drdp \quad (\text{Eq. \eqref{eqn: psi0}} )\nonumber \\
    &= -  \partial_t \big[ \Psi_0 \langle\chi \rangle_\pi \big] - \gamma \partial_\phi \big[ \Psi_0 \langle\chi (p - \phi) \rangle_\pi \big]. \label{eqn: integral computation}
\end{align}
Recalling Eq. \eqref{eqn: pi moments}, we compute
\begin{align}
    &\langle \chi \rangle_\pi = a \bigg(  \frac{1}{\lambda_r} +\frac{1}{\lambda_p} \bigg), & &\langle \chi (p - \phi) \rangle_\pi =  \langle \chi \rangle_\pi (a-\phi) + \frac{\text{Var}_r}{\lambda_r},
\end{align}
and so
\begin{align}
    \partial_t \big[ \Psi_0 \langle\chi \rangle_\pi \big] &=  \partial_t \Psi_0 \langle \chi \rangle_\pi + \Psi_0 \partial_t \langle \chi \rangle_\pi \nonumber \\
    &=-\gamma \partial_\phi \big[ (a - \phi) \Psi_0 \big] + \Psi_0 \partial_t \langle \chi \rangle_\pi \quad \text{(Eq. \eqref{eqn: f0 result})} \nonumber \\
    &= -\gamma \partial_\phi \big[ (a - \phi) \Psi_0 \big] \langle \chi \rangle_\pi (a-\phi) +  \partial_t a \bigg( \frac{1}{\lambda_r} + \frac{1}{\lambda_p} \bigg) \Psi_0,
\end{align}
and 
\begin{align}
    \gamma \partial_\phi \big[ \Psi_0 \langle\chi (p - \phi) \rangle_\pi \big] &= \gamma \partial_\phi \big[(a-\phi) \Psi_0  \big] \langle \chi \rangle_\pi + \frac{\gamma \text{Var}_r}{\lambda_r} \partial_\phi \Psi_0.
\end{align}
Hence, Eq. \eqref{eqn: integral computation} simplifies to
\begin{align}
     \int_0^\infty \int_0^\infty (p-a) \psi_1 drdp &=  -\partial_t a \bigg( \frac{1}{\lambda_r} + \frac{1}{\lambda_p} \bigg) \Psi_0 - \bigg(\frac{\gamma \, \text{Var}_r}{\lambda_r} \bigg) \partial_\phi \Psi_0 
\end{align}
and we obtain
\begin{align}
    \int_0^\infty \int_0^\infty \mathcal{L}_1 \psi_1 drdp  &= \gamma \partial_\phi \bigg[ (a - \phi) \Psi_1 - \partial_t a \bigg( \frac{1}{\lambda_r} + \frac{1}{\lambda_p} \bigg) \Psi_0 - \bigg(\frac{\gamma \, \text{Var}_r}{\lambda_r} \bigg) \partial_\phi \Psi_0\bigg].
\end{align}
Finally, returning to Eq. \eqref{eqn: f1 pde} reveals that the dynamics of $\Psi_1$ are given by
\begin{align}
    \partial_t \Psi_1 + \gamma \partial_\phi \bigg[ (a - \phi) \Psi_1 - \partial_t a \bigg( \frac{1}{\lambda_r} + \frac{1}{\lambda_p} \bigg) \Psi_0 - \bigg(\frac{\gamma \, \text{Var}_r}{\lambda_r} \bigg) \partial_\phi \Psi_0\bigg] = 0. \label{eqn: f1 result}
\end{align}
Recalling that $\Psi = \Psi_0 + \epsilon \Psi_1 + \dots$, we sum Eq. \eqref{eqn: f0 result} and $\epsilon \times$ Eq. \eqref{eqn: f1 result} to find that the dynamics of $\Psi$ up to $\mathcal{O}(\epsilon)$ contributions is the advection-diffusion equation \eqref{eqn: phenotype effective}.

\clearpage
\section{Early population dynamics} \label{Appendix: early population dynamics}

\begin{figure}[h]
    \centering
    \includegraphics[width=0.99\textwidth]{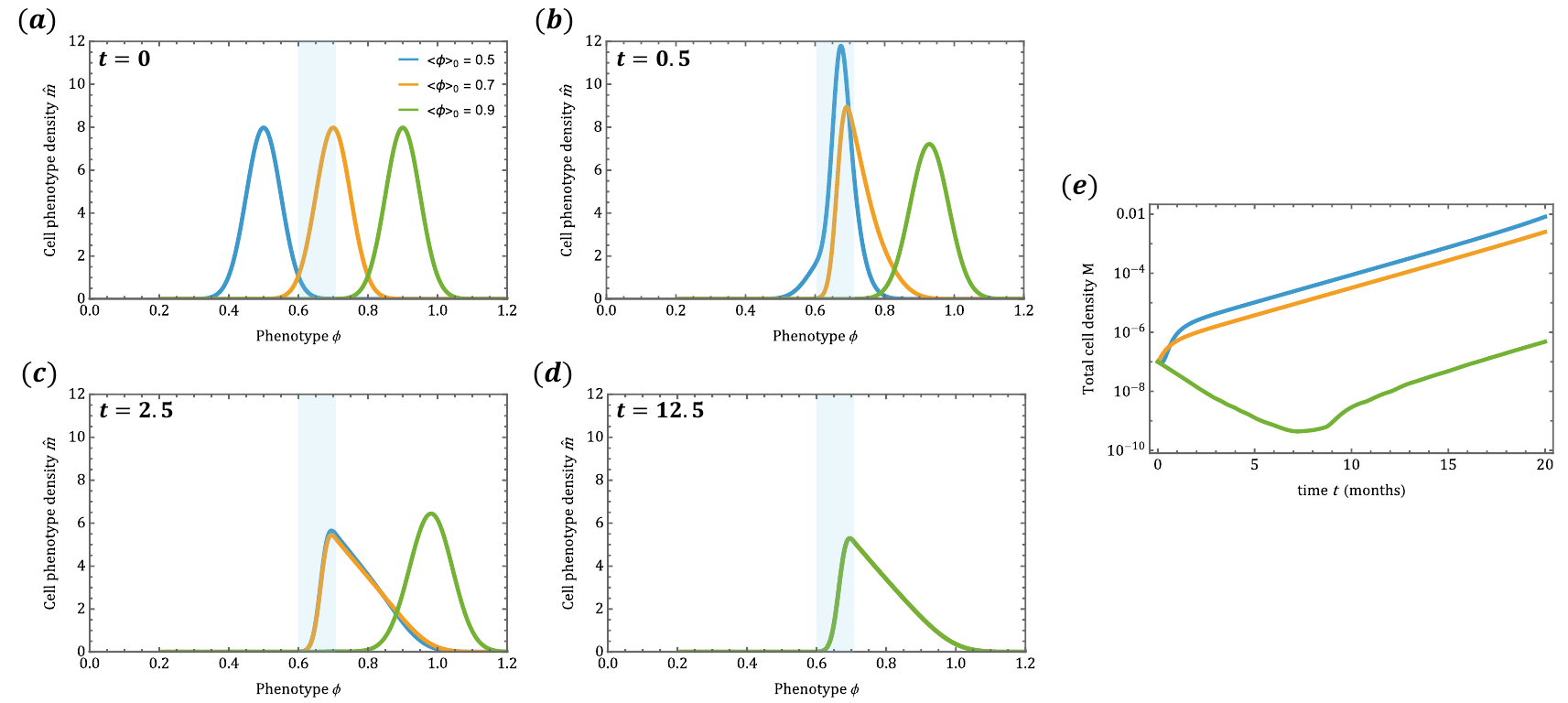}
    \caption{\textbf{Transients from independent initial conditions merge into a common exponential growth profile. } The depicted initial phenotype distributions are Gaussian and $M_0 = 10^{-7}$. Parameter values are set to: $\phi_L = 0.60$, $\phi_R = 0.71$, $\rho_\text{max} = 10.4$, $\nu = 0.97$, $\gamma = 0.67$, $q = 1.79$, $\theta = 0.19$, $\lambda_r = 0.28$, $\lambda_p = 0.35$ and $\kappa = 2$.}
    \label{fig: early dynamics}
\end{figure}

\clearpage
\section{Xenograft invasive-to-proliferative switching times} \label{Appendix: Hoek} 

Hoek \textit{et al.} \cite{hoek2008vivo}  present xenograft experiments in which human melanoma cell lines were transplanted into immunocompromised mice. They found the proliferative cell lines initiated tumour growth (volume $>100 \, $mm$^3$) in $14 \pm 3$ days, while invasive cell lines, which expressed low or undetectable levels of MITF in culture, required $59 \pm 11$ days to reach the same size. Tumours were indistinguishable upon excision.

The authors interpret the lag in tumour initiation times as being due to the time required for invasive to proliferative phenotype switching. 
Following the authors' interpretation, we model the tumour initiation times in terms of random variables, $\eta_1, \eta_2 > 0$:
    \begin{align}
        &T_\text{pro}^{\text{obs}, i} = \eta_1, & &T_\text{inv}^{\text{obs}, i} = \eta_1 + \eta_2.
    \end{align}
Here $\eta_1$ is the time taken for proliferative-seeded tumours to attain the initiation volume ($>100 \, $mm$^3$), which we assume follows a lognormal distribution
    \begin{align}
        \eta_1 \sim \text{Lognormal}(\log(\mu_1), \sigma_1^2).
    \end{align}
This is reasonable since the proliferative-seeded tumours are expected to follow exponential growth rapidly after seeding. 
The random variable $\eta_2$ represents the time for the invasive-seeded population to undergo proliferative phenotype switching; we expect this time to vary between measurements due to microenvironment variation between mice.
Given that the invasive cells expressed low or undetectable levels of MITF prior to seeding, for consistency with our mathematical model we expect $\eta_2$ to be distributed about the value
\begin{align}
    t_s = - \gamma^{-1} \log (1 - \phi_L). \label{eqn: ts app}
\end{align}
Eq. \eqref{eqn: ts app} is the time required for cells to drift from $\phi = 0$ into the proliferative window $\phi \in (\phi_L, \phi_R)$ by advection under minimal stress conditions. Hence, we model $\eta_2$ as  
\begin{align}
    \eta_2 \sim \text{Lognormal}(\log(t_s), \sigma_2^2),
\end{align}
where it follows that $\text{median}(\eta_2) = t_s$. 
The parameter $\sigma_2$ accounts for the microenvironment variation between mice.

The corresponding log-likelihood of the experiment observations is
\begin{align}
    \mathcal{L}_\text{xenograft} &= \mathcal{L}_\text{pro} + \mathcal{L}_\text{inv}, \nonumber \\
    \mathcal{L}_\text{pro} &= \sum_i \bigg( - \log \big[ 2 \pi \sigma_1 T^{\text{obs}, i}_\text{pro} \big] \nonumber \\
    &\qquad \qquad + \frac{1}{2 \sigma_1^2} \big[ \log(\mu_1) - \log (T^{\text{obs}, i}_\text{pro}) \big]^2 \bigg), \nonumber \\
    \mathcal{L}_\text{inv} &= \sum_i \log \int_0^{T_\text{inv}^{\text{obs}, i}} f_{1}(T) \cdot f_2(T_\text{inv}^{\text{obs}, i}-T) \, dT, 
\end{align}
where 
\begin{align}
    &f_1(T) = \frac{1}{\sqrt{2 \pi} \sigma_1 T} \exp \bigg( - \frac{(\log T - \log \mu_1)^2}{2 \sigma_1^2} \bigg), \\ &f_2(T) = \frac{1}{\sqrt{2 \pi} \sigma_2 T} \exp \bigg( - \frac{(\log T - \log t_s)^2}{2 \sigma_2^2} \bigg).
\end{align}
Note that the probability density function of the sum of lognormal random variables has no closed form and so the convolution must be computed explicitly. 
The above likelihood function and uniform positive priors on $\mu_1$, $\sigma_1$, $t_s$ and $\sigma_2$ constitute a closed inference problem that we use to estimate $t_s$. 

\begin{figure}
    \centering
    \includegraphics[width=0.99\textwidth]{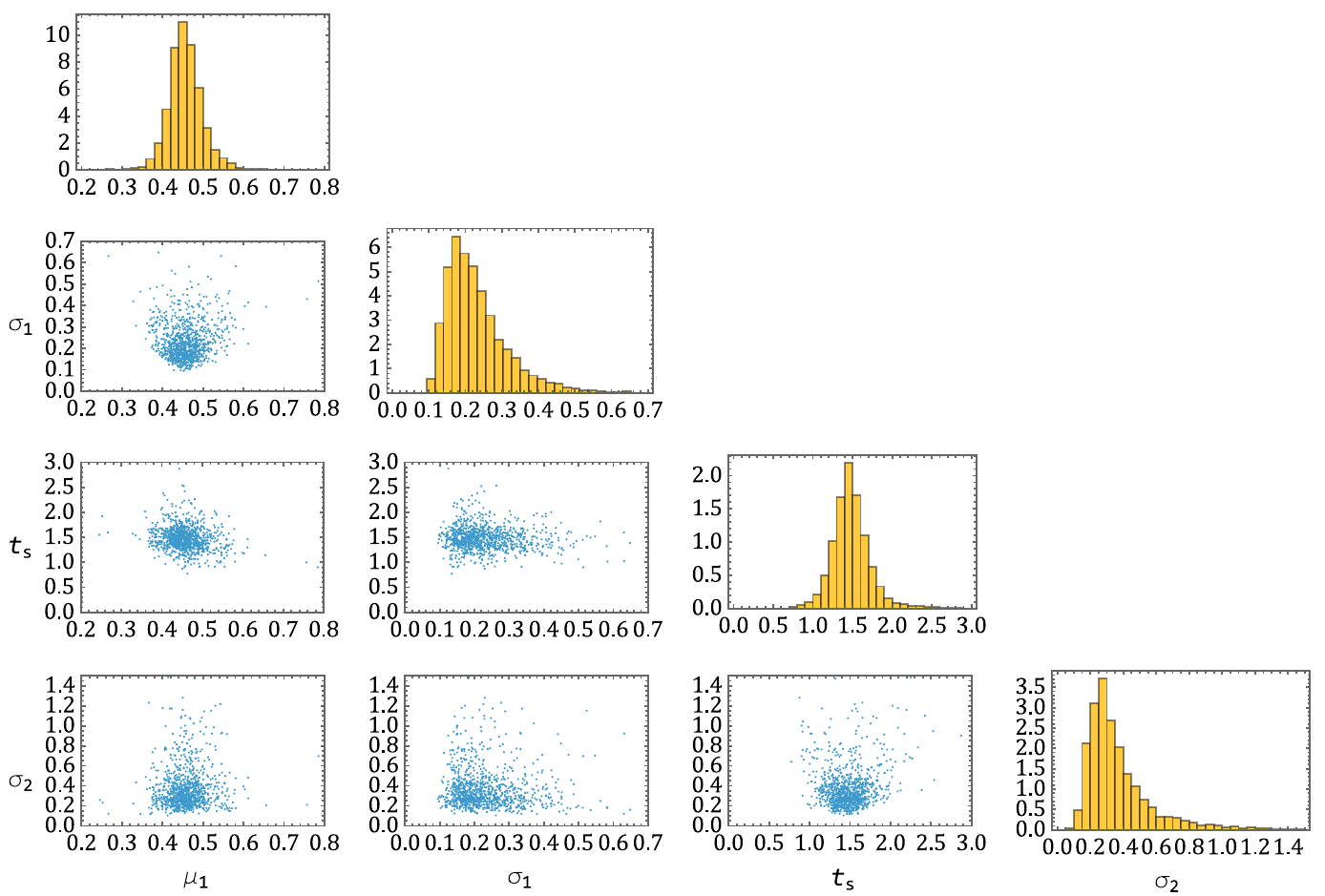}
    \caption{\textbf{Posterior samples for the parameters in the Xenograft inference sub-problem. } The plots show samples from 3 independent chains of length $5 \times 10^4$ (including $10^4$ burn-in samples). We note that all parameters are practically identifiable. The second-order interpolation of the histogram for $t_s$ defines the prior distribution for $t_s$ in Sect. \ref{sec: population model parameter calibration}. }
    \label{fig: Hoek_pairwise}
\end{figure}

\begin{figure}
    \centering
    \includegraphics[width=0.75\textwidth]{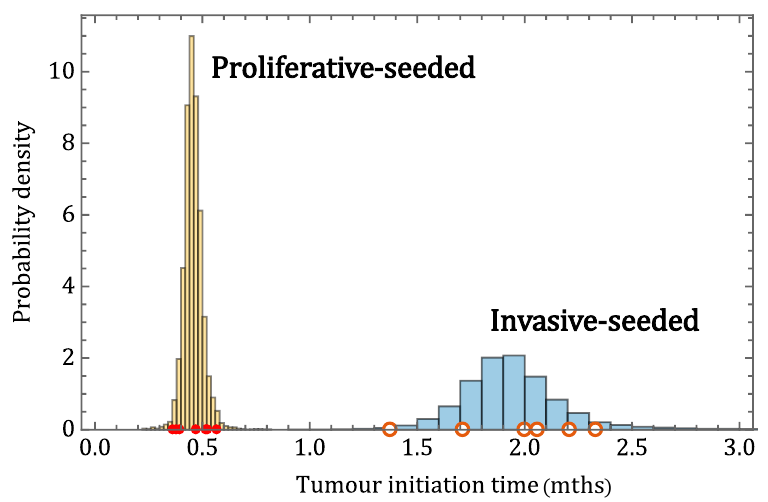}
    \caption{\textbf{Posterior predictions for the Xenograft inference sub-problem. } The left histogram represents samples of $\mu_1$ and the right histogram represents samples of $\mu_1 + t_s$. The samples cover the data well. }
    \label{fig: Hoek_posterior_predictions}
\end{figure}

Posterior distribution samples are supplied in Fig. \ref{fig: Hoek_pairwise} and their predictions verified against the data in Fig. \ref{fig: Hoek_posterior_predictions}. The second-order interpolation of the histogram for $t_s$ defines its prior distribution in Sect. \ref{sec: population model parameter calibration}.

\clearpage
\section{TUNEL index} \label{Appendix: TUNEL}
Shukuwa \textit{et al.} \cite{shukuwa2002fas} present clinical data for the TUNEL index derived from cutaneous melanoma. The measurement represents the proportion of dead cells in the sample. We label the data as $\mathcal{T}_i$ and assume a lognormal observation model 
\begin{align}
    &\ln(\mathcal{T}_i) = \ln(\mathcal{T}_\text{model}) + \eta, & &\eta \sim \text{Normal}(0, \sigma^2). \label{eqn: TUNEL obs}
\end{align}
We derive an expression for $\mathcal{T}_\text{model}$ by considering the following equation for TUNEL-positive (dead) cell density
\begin{align}
    \frac{d N}{dt} = \nu M - \delta_N N, \label{eqn: N}
\end{align}
where $\delta_N$ is the loss rate of TUNEL-positive dead cells due to either efferocytosis or degradation of TUNEL itself. Since efferocytosis is expected to be highly efficient, with typical estimates on the order of minutes to hours \cite{morioka2019living}, we let $\delta_N = \epsilon^{-1} \hat{\delta}_N$ so that the TUNEL loss rate is measured in hours. Assuming the death rate $\nu$ occurs on the months timescale of population dynamics, Eq. \eqref{eqn: N} admits the following quasi-steady expression for the TUNEL index as $\epsilon \rightarrow 0$:
\begin{align}
    &\mathcal{T}_\text{model} := \frac{N}{N+M} \sim \frac{\epsilon R}{1+ \epsilon R}, & &R := \frac{\nu}{\hat{\delta}_N} = \mathcal{O}(1).
\end{align}
Recalling that $\epsilon = \frac{1}{720}$ is the ratio of an hour to a month, $R$ represents the ratio of the death rate in months to the TUNEL loss rate in hours. 

\begin{figure}[h]
    \centering
    \includegraphics[width=0.75\textwidth]{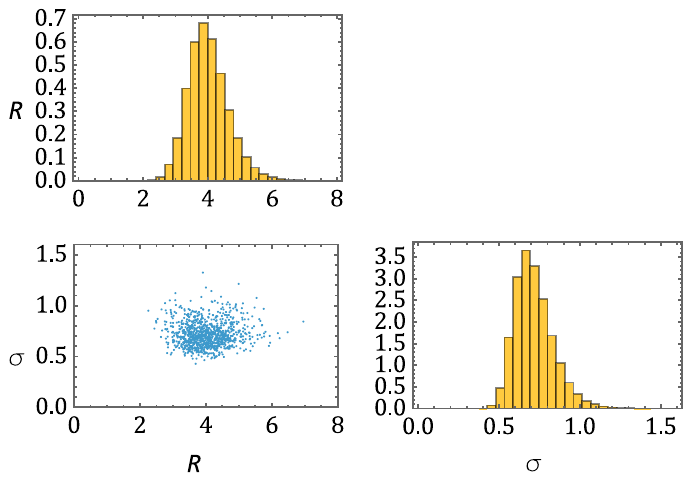}
    \caption{\textbf{Posterior samples for parameters of the TUNEL inference sub-problem. } The plots show samples from 3 independent chains of length $5 \times 10^4$ (including $10^4$ burn-in samples). }
    \label{fig: TUNEL_pairwise}
\end{figure}

\begin{figure}
    \centering
    \includegraphics[width=0.99\textwidth]{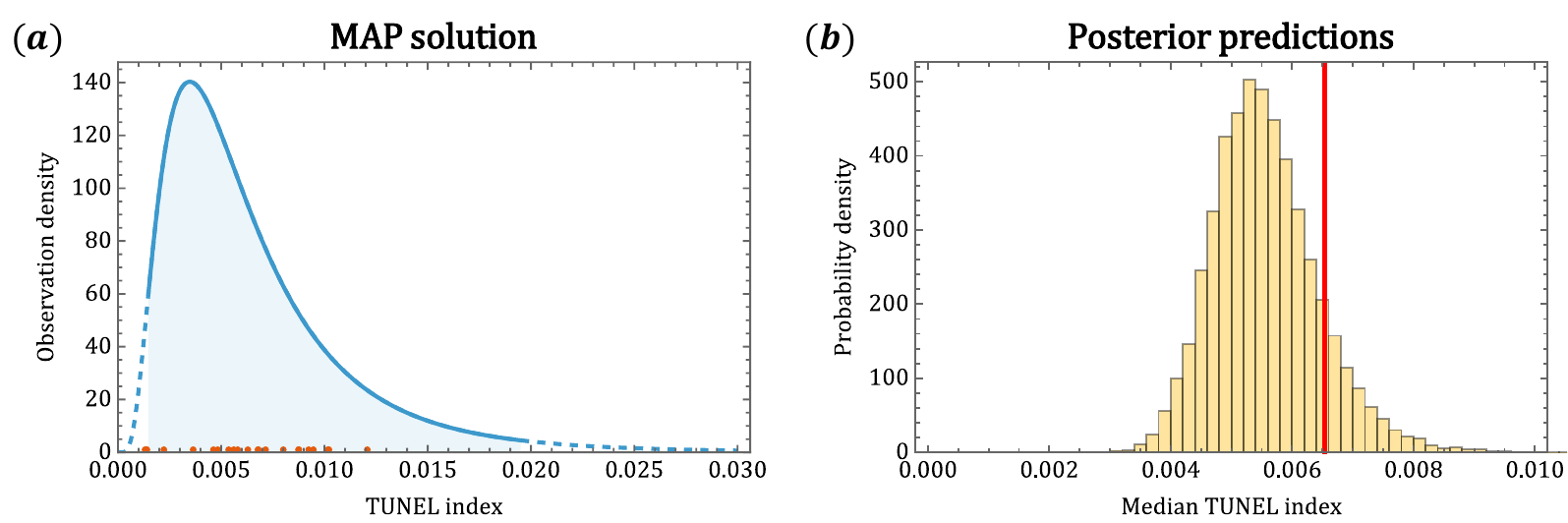}
    \caption{\textbf{Diagnostic plots for the TUNEL inference sub-problem. } Plot (a) shows the best-fit (MAP) solution $R = 3.89$, $\sigma = 0.66$. The 95$\%$ observation window is shaded and covers the data well. Plot (b) shows posterior predictions of the median TUNEL measurement with the data median indicated via the red line. We note that the data median is within the range of the simulated medians. }
    \label{fig: TUNEL_diagnostics}
\end{figure}

We use Bayesian inference to infer the value $R$ according to the observation model \eqref{eqn: TUNEL obs}. We assume uniform positive priors on $R$ and $\sigma$. Posterior samples are shown in Fig. \ref{fig: TUNEL_pairwise}. The results are localised bell-curves that suggest practical identifiability. Diagnostic plots are provided in Fig. \ref{fig: TUNEL_diagnostics}. Plot (a) shows that the MAP solution provides good coverage of the data. Plot (b) indicates that the data median is within the range of the medians predicted from the posterior samples. 

Samples for $\nu = R \cdot \hat{\delta}_N$ may be obtained by specifying a distribution of values for the TUNEL loss rate $\hat{\delta}_N$. Direct information on dead cell clearance rates in the skin is limited. However, a mathematical model of psoriasis indicates that apoptosis and subsequent uptake occurs in the epidermis within 6 hours \cite{weatherhead2011keratinocyte}. We note that the engulfment process itself typically takes 10 minutes or slightly longer \cite{taruc2018quantification}. Hence, it is reasonable to assume that $10 \text{ min} < \hat{\delta}_N^{-1} < 6 \,\text{h}$. In the absence of further information, we apply uniform uncertainty in the uptake time and, assuming independence between $R$ and $\hat{\delta}_N$, generate samples for $\nu$ according to
\begin{align}
    &\nu = \frac{R}{\hat{\delta}^{-1}_N}, & &R \sim \text{posterior distribution}, & &\hat{\delta}^{-1}_N \sim \text{Uniform}\bigg(\frac{1}{6}, 6\bigg). \label{eqn: nu prior samples}
\end{align}
A kernel density estimate of the samples from Eq. \eqref{eqn: nu prior samples} defines the prior distribution for $\nu$ that we use in Sect. \ref{sec: population model parameter calibration}. 

\clearpage
\section{MCMC plots: population inference problem} \label{Appendix: population_MCMC}

\begin{figure}[h]
    \centering
    \includegraphics[width=0.99\textwidth]{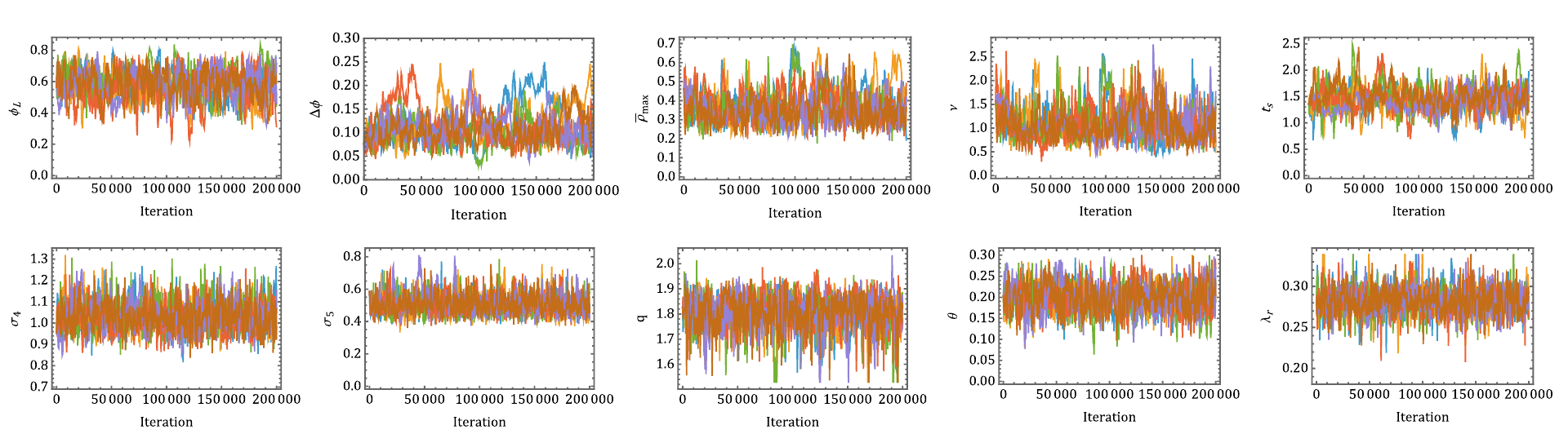}
    \caption{\textbf{Trace plots of the six independent chains.} The chains mix well and remain stationary. The Gelman-Rubin statistic is less than $1.01$ for each parameter. Each chain consists of $2 \times 10^6$ iterations non burn-in iterations, here thinned by a factor 10 for visualisation.}
    \label{fig: population_chains}
\end{figure}

\begin{figure}
    \centering
    \includegraphics[width=0.99\textwidth]{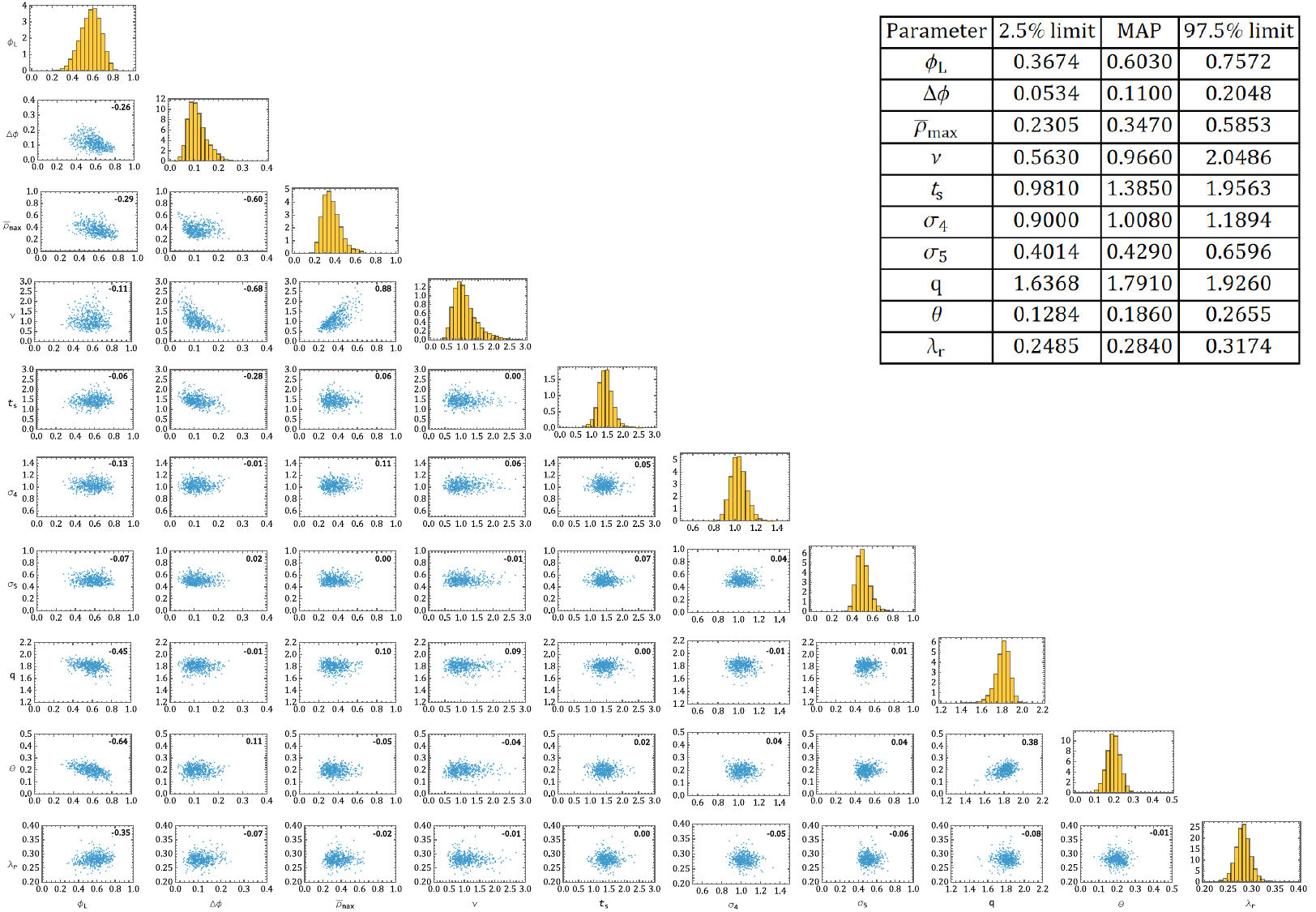}
    \caption{\textbf{Posterior samples for all parameters in the population inference problem. } Samples for each parameter are localised and form a smooth bell-curve histogram. The distributions of the subcellular parameters $(q, \theta, \lambda_r)$ are largely the same as that in Fig. \ref{fig: subcellular_MCMC_full}, including the positive $(q, \theta)$ correlation. Pearson correlation coefficients are indicated in bold in the pairwise plots. The table shows the MAP values and bounds of the 95\% credible interval for each parameter. }
    \label{fig: population_pairwise_full}
\end{figure}

\begin{figure}
    \centering
    \includegraphics[width=0.99\textwidth]{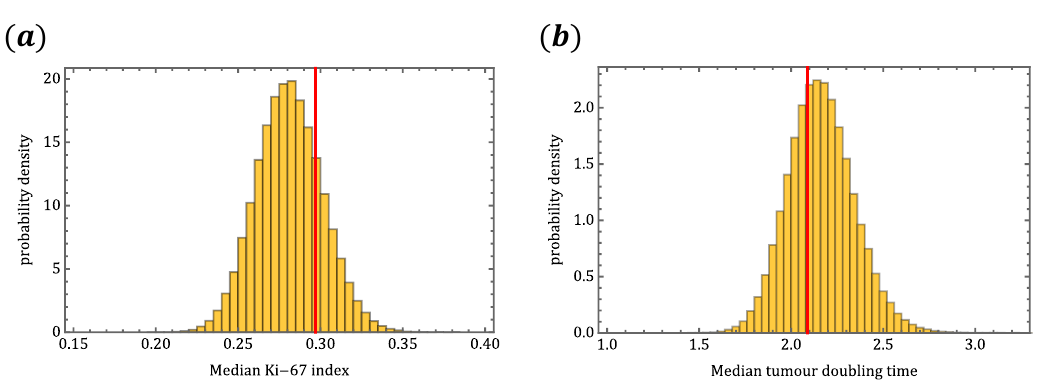}
    \caption{\textbf{Posterior predictions for the observation distribution medians cover the data medians. } Plots (a) and (b) are histograms of the posterior predictions for the Ki-67 and tumour doubling time medians, respectively, with data medians overlaid as a red line. }
    \label{fig: population_posterior_predictions}
\end{figure}

\begin{figure}
    \centering
    \includegraphics[width=0.8\textwidth]{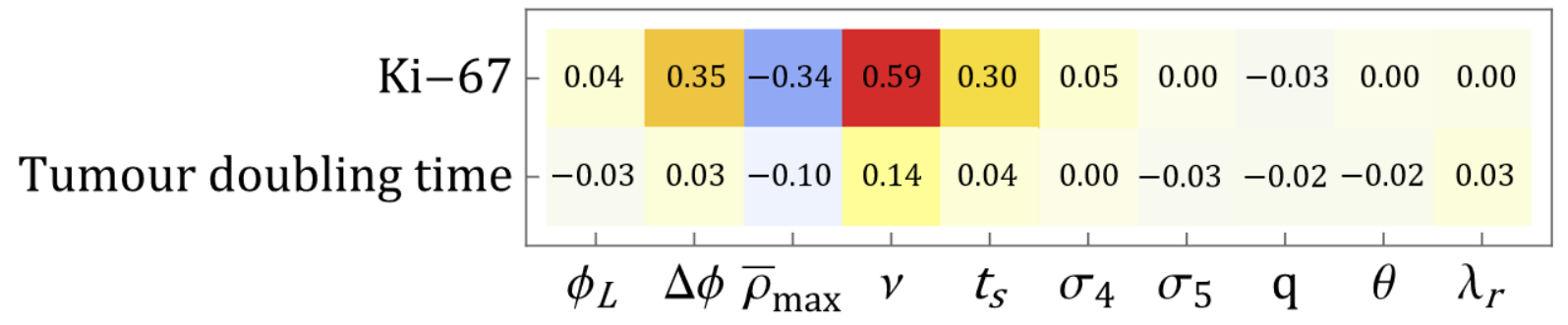}
    \caption{\textbf{Posterior predictive sensitivity analysis. } The values indicate the Pearson Correlation Coefficient for each of the inference parameters against the two outputs. }
    \label{fig: sensitivity}
\end{figure}

\clearpage
\section{Details of the spatially resolved model} \label{Appendix: spatial}

In the spatially resolved population model of Sect. \ref{sec: spatial}, the phenotype drift and diffusivity are
\begin{align}
    &v(\phi, r, t) := \gamma \bigg[(a - \phi) - \epsilon \frac{\partial a}{\partial t} \bigg( \frac{1}{\lambda_r} + \frac{1}{\lambda_p}  \bigg) \bigg], & &\hat{D}(r,t) := \frac{\epsilon \gamma^2\text{Var}_r(a)}{\lambda_r},
\end{align}
where the average transcription rate is coupled to the local cell density
\begin{align}
    &a(r,t) = \frac{1}{1 + M(r,t)/M^\star}, & &M(r,t)  = \int_0^\infty m(\phi, r, t) d\phi.
\end{align}
The proliferation rate 
\begin{align}
    \rho(\phi, r, t) = \frac{\rho_\text{max}}{1 + \kappa M(r,t)/M^\star} \cdot W(\phi; \phi_L, \phi_R)
\end{align}
also acquires spatial dependence from $M$. \\


\noindent \textbf{Explicit rates of change. } Integrating the PDE \eqref{eqn: m spatial} with respect to $\phi$ and enforcing the boundary conditions \eqref{eqn: m spatial noflux} yields
\begin{align}
    \frac{\partial M}{\partial t} = \int_0^\infty \big[ D(\phi) \nabla^2m + \big(\rho (\phi, r,t) - \nu \big)m \big] d\phi,
\end{align}
where the first term in the integrand accounts for local density change due to phenotype-dependent cell movement. 
Hence the rate of change for $a(r,t)$ is
\begin{align}
    \frac{\partial a}{\partial t} = - \frac{\frac{\partial M}{\partial t}}{M^\star (1 + M/M^\star)^2} = - \frac{\int_0^\infty \big[ D(\phi) \nabla^2m + \big(\rho (\phi, r,t) - \nu \big)m \big] d\phi }{M^\star (1 + M/M^\star)^2}. \label{eqn: dadt spatial}
\end{align}

\noindent \textbf{Non-dimensionalisation. } We scale the model as follows
\begin{align}
    &\tilde{m}(\phi, \tilde{r},t) = \frac{m(\phi, r, t)}{M^\star}, & &\tilde{M}(t) = \frac{M(t)}{M^\star}, & &\tilde{r} = \frac{r}{[1 \text{mm}]}.
\end{align}
Consistent with Sect. \ref{sec: population}, cell densities are measured with respect to $M^\star$, the density for which MITF transcription is half-maximal. 
We measure the spatial coordinate in millimetres. 


\end{document}